\documentclass[preprint,journal]{vgtc}            

\onlineid{0}

\preprinttext{Authors’ version. To appear in IEEE Transactions on Visualization and Computer Graphics.}

\vgtccategory{Research}

\vgtcpapertype{please specify}

\title{See What I Mean? Mobile Eye-Perspective Rendering for Optical See-through Head-mounted Displays}

\author{%
 \authororcid{Gerlinde Emsenhuber}{0000-0001-7776-586X}, 
 \authororcid{Tobias Langlotz}{0000-0003-1275-2026}, 
 \authororcid{Denis Kalkofen}{0000-0002-0359-206X}, and
 \authororcid{Markus Tatzgern}{0000-0002-3900-4944}
}

\authorfooter{
 \item
 	Gerlinde Emsenhuber is with Salzburg University of Applied Sciences, Austria and Graz University of Technology, Austria.
 	E-mail: gerlinde.emsenhuber@fh-salzburg.ac.at.
 \item
 	Tobias Langlotz is with Aarhus University, Denmark.
 	E-mail: tobias.langlotz@cs.au.dk.
 \item Denis Kalkofen is with Graz University of Technology, Austria.
 	E-mail: kalkofen@tugraz.at.
\item Markus Tatzgern is with Salzburg University of Applied Sciences, Austria.
  	E-mail: markus.tatzgern@fh-salzburg.ac.at.  
}

\abstract{
Image-based scene understanding allows Augmented Reality (AR) systems to provide contextual visual guidance in unprepared, real‑world environments. While effective on video see‑through (VST) head‑mounted displays (HMDs), such methods suffer on optical see‑through (OST) HMDs due to misregistration between the world‑facing camera and the user’s eye perspective. To approximate the user's true eye view, we implement and evaluate three software‑based eye‑perspective rendering (EPR) techniques on a commercially available, untethered OST HMD (Microsoft HoloLens~2): (1) Plane‑Proxy EPR, projecting onto a fixed‑distance plane; (2) Mesh‑Proxy EPR, using SLAM‑based reconstruction for projection; and (3) Gaze‑Proxy EPR, a novel eye‑tracking‑based method that aligns the projection with the user’s gaze depth. A user study on real‑world tasks underscores the importance of accurate EPR and demonstrates gaze‑proxy as a lightweight alternative to geometry‑based methods. We release our EPR framework as open source.
}


\teaser{
  \centering
  \includegraphics[width=\linewidth]{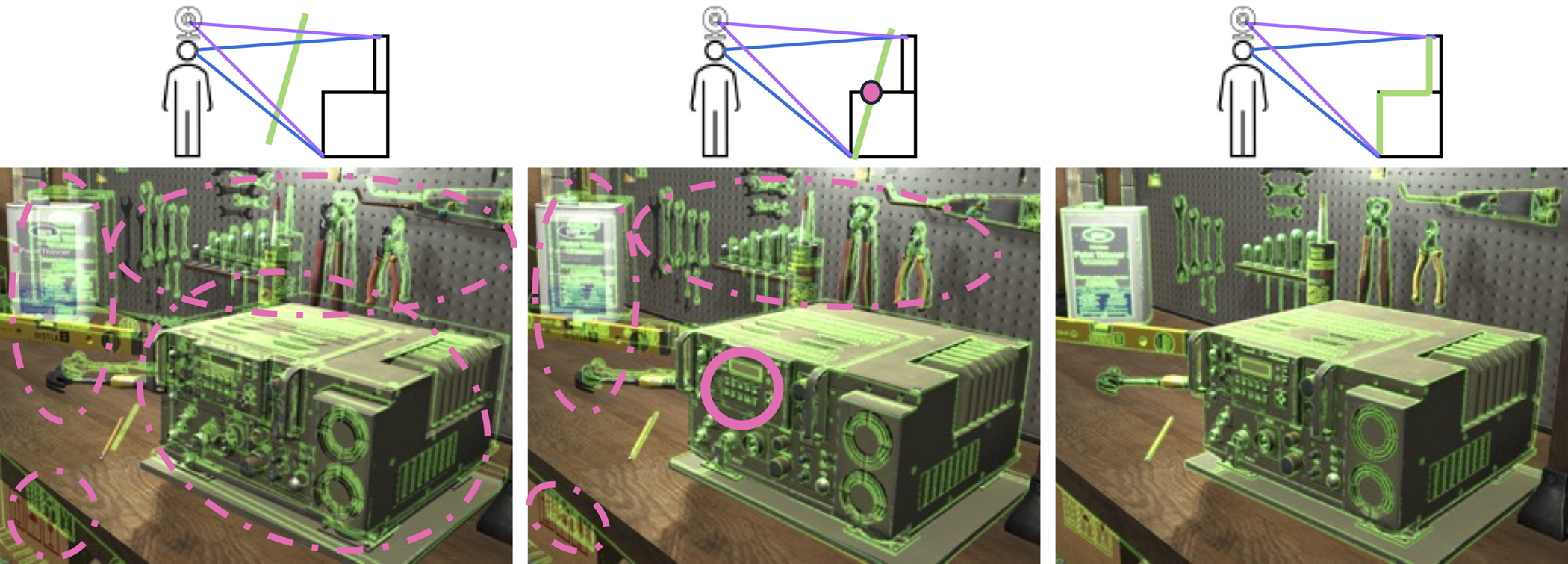}
\caption{
Edges are extracted from a head‑mounted display (HMD) camera view (purple viewport), e.g., to support people with low vision by highlighting scene elements. For accurate alignment, the edges must be shown from the user’s eye perspective (blue viewport) on the optical see‑through (O ST) display. (Left) A naive \eprPlane{} method projects the camera view onto a fixed‑distance proxy plane (green), causing misalignment between visualization and real‑world objects (dashed pink areas). (Middle) Our novel \eprGaze{} approach uses eye‑tracking to dynamically adjust the plane to match the depth at the user’s gaze point (center pink dot/circle), improving alignment at the focal area but causing misalignment elsewhere. (Right) A \eprMesh{} technique projects the camera view onto a scene reconstruction (green), ensuring accurate alignment from all perspectives.
} 
  \label{fig:teaser}
}

\graphicspath{{figs/}{figures/}{pictures/}{images/}{./}} 

\usepackage{tabu}                      
\usepackage{booktabs}                  
\usepackage{lipsum}                    
\usepackage{mwe}                       
\usepackage[normalem]{ulem}            

\usepackage{mathptmx}                  
\usepackage{acronym}
\usepackage{tabularx}
\usepackage[table,svgnames]{xcolor}
\usepackage{balance}

\newcommand{\art}[4]{$F$(#1,#2)$=$#3,p#4} 
\newcommand{\artcon}[3]{t(#1)$=$#2,p#3}
\newcommand{\arttext}[4]{~(\art{#1}{#2}{#3}{#4})}
\newcommand{\artcontext}[4]{~(\artcon{#1}{#2}{#3}{#4})}

\newcommand{\condWall}{WALL}
\newcommand{\condTable}{TA\-BLE}
\newcommand{\condMesh}{MESH}
\newcommand{\condFixed}{PLA\-NE}
\newcommand{\condGaze}{GA\-ZE}
\newcommand{\eprPlane}{Plane-Proxy EPR}
\newcommand{\eprGaze}{Gaze-Proxy EPR}
\newcommand{\eprMesh}{Mesh-Proxy EPR}

\acrodef{UI}[UI]{User Interface}
\acrodef{XR}[XR]{Extended Reality}
\acrodef{AR}[AR]{Augmented Reality}
\acrodef{VR}[VR]{Virtual Reality}
\acrodef{MR}[MR]{Mixed Reality}
\acrodef{OST}[OST]{optical see-through}
\acrodef{VST}[VST]{video see-through}
\acrodef{HMD}[HMD]{head-mounted display}
\acrodefplural{HMD}[HMDs]{head-mounted displays}
\acrodef{IBR}[IBR]{image-based rendering}
\acrodef{TCT}[TCT]{Task Completion Time}
\acrodef{SEQ}[SEQ]{Single Ease Question}
\acrodef{SUS}[SUS]{System Usability Scale}
\acrodef{XRP}[XRP]{Extended Reality Prototyping}
\acrodef{UX}[UX]{user experience}
\acrodef{TCT}[TCT]{task completion time}
\acrodef{EPR}[EPR]{eye-perspective rendering}
\acrodef{IMU}[IMU]{inertial measurement unit}
\acrodef{CVD}[CVD]{color vision deficiency}
\acrodef{DNN}[DNN]{deep neural network}
\acrodef{VLM}[VLM]{vision language model}
\acrodefplural{VLM}[VLMs]{vision language models}
\acrodef{LLM}[LLM]{large language model}
\acrodefplural{LLM}[LLMs]{large language models}
\acrodef{AI}[AI]{Artificial Intelligence}
\acrodef{MRTK}[MRTK]{Mixed Reality Toolkit}

\begin{document}


\firstsection{Introduction}

\maketitle

\ac{AR} combined with \ac{OST} \acp{HMD} overlays virtual information onto the user's view of the real world. Such systems support assembly and repair tasks~\cite{Eswaran2023ARAssembly}, assist first responders in hazardous scenarios~\cite{Bhattarai2020Firefighting, Liu2024HazardDetection}, and aid elderly and visually impaired users~\cite{Htike2021, Younis2019, Yu2018}. Commonly, \ac{AR} visualizations highlight task-relevant objects identified through the analysis of images captured by the built-in, world-facing camera of the \ac{HMD}. Traditionally, these highlighted objects rely on manually preregistered semantic annotations or automatic detection by specialized computer vision models, restricting support to predefined object types and relationships. 

The emergence of context-aware \ac{AI}, such as \acp{VLM}, \acp{LLM}, and open-vocabulary scene graphs~\cite{Gu2024, Xu2024}, reduces the need for extensive prior annotation by enriching captured imagery with semantic context. Image-based scene understanding facilitates a broader adoption of \ac{AR} in mobile, unprepared environments~\cite{10.1145/3654777.3676379} that can be directly applied to \ac{VST} \ac{HMD} setups where the world-facing cameras replace the user's eyes. However, when using an \ac{OST} device, mapping the image information extracted from a world-camera view to the user's eyes remains challenging due to their different perspectives (\cref{fig:teaser}(Left)). 

Previous work has developed \ac{HMD} prototypes integrating cameras that capture the user's exact viewpoint by redirecting light through the transparent display using half-silvered mirrors~\cite{Langlotz2016, Langlotz2018}. This setup allows direct usage of the captured image for highlighting object silhouettes~\cite{Mooser2007} or adjusting scene colors to assist individuals with a \ac{CVD}~\cite{Langlotz2018}. However, these hardware modifications cannot be applied to existing commercially available \acp{HMD}, limiting immediate applicability while also significantly increasing the size of the \acp{HMD}. 

Software-based alternatives use scene information to reconstruct the user's perspective through \ac{EPR} techniques. The most straightforward method, which we refer to as \eprPlane{}, projects the analyzed world-camera view onto a virtual plane placed at a fixed distance in front of the user~\cite{Htike2021,Zhu2023}. This approach achieves alignment when the virtual plane is aligned in depth with the real-world geometry, but results in visual misalignment when there is an offset (\cref{fig:teaser}(Left)).

A more accurate method addresses this by projecting the world-camera view onto a 3D mesh reconstructed from depth maps or SLAM data~\cite{10.1145/3528233.3530701, 10.1145/3384540, Emsenhuber2023}. This method, which we refer to as \eprMesh{}, significantly improves the alignment~\cite{Emsenhuber2022} between image-based overlays and real-world geometry (\cref{fig:teaser}(Right)), but requires a full 3D reconstruction, which may be computationally expensive, battery-draining, and not always available in real-time.

We further propose the \eprGaze{} method, a novel approach that dynamically aligns the projection plane of \eprPlane{} with the scene depth at the user’s current gaze target. Unlike \eprMesh{}, which requires a 3D reconstruction of the environment, \eprGaze{} instead uses real-time eye tracking to estimate fixation depth, enabling localized alignment without the need for spatial meshes. While \eprMesh{} aims for global alignment across the entire field of view, \eprGaze{} prioritizes accuracy in the user’s focus region (\cref{fig:teaser}(Middle)), adapting the projection plane as the gaze shifts. This makes it particularly suitable for mobile \ac{OST} \ac{HMD}s, as it relies solely on eye vergence for depth estimation when precise eye tracking is available~\cite{Mardanbegi2019}, whereas scene reconstructions may be unavailable, unreliable, or too slow to update (e.g., 1-5 FPS for HoloLens~2).
 
In this paper, we explore software-based \ac{EPR} for mobile, untethered \ac{OST} \acp{HMD} by implementing the three aforementioned \ac{EPR} approaches on a commercially available hardware (HoloLens~2). Previous studies have demonstrated the value of \ac{EPR} for adapting virtual content to avoid interference with the real-world background~\cite{Langlotz2016, Emsenhuber2023} and assisting individuals with \ac{CVD}~\cite{Langlotz2018} in tethered, sedentary setups that restrict user mobility and do not allow users to interact with the real world. In contrast, we demonstrate the effectiveness of \ac{EPR} in a mobile, untethered scenario. We also require users to interact with the augmented real-world scene, thereby, underlining the need for \ac{EPR} to enable accurate interactions. By providing our \ac{EPR} framework as open-source software, we facilitate further research beyond controlled laboratory settings and simulations conducted in \ac{VR}~\cite{Burova2020, Cheng2021}, addressing the lack of suitable \ac{EPR} frameworks for untethered \ac{OST} \acp{HMD}.
 
In a user study, we compare the three \ac{EPR} methods in terms of interaction accuracy, usability, and task load. By also collecting qualitative feedback, we provide quantitative and qualitative insights into the use of \ac{EPR} using an untethered device. 

In summary, we make the following contributions: 
\begin{itemize} 
\item We introduce a novel \eprGaze{} method for \ac{OST} \acp{HMD} that uses eye tracking to align the projection plane with the scene depth at the user's current focus, providing localized accuracy with minimal computational overhead.
\item We implement and compare all three \ac{EPR} methods on a state-of-the-art, untethered \ac{OST} \ac{HMD} (HoloLens~2), providing quantitative and qualitative insights from the first exploration of \ac{EPR} effects during interactive tasks involving real-world environments.
\item Our results clearly demonstrate the need for accurate \ac{EPR} for image-based analysis, highlighting the impact of misalignments between the world camera and the user’s eyes to raise awareness of this challenge.
\item We release all three \ac{EPR} implementations and the study framework as open source, facilitating future research and development of contextual, image-based \ac{AR} support systems for mobile, untethered use cases. 
\end{itemize}

Please note that the examples used to illustrate \ac{EPR} techniques in this paper are generated within synthetic scenes. Capturing \ac{EPR}-aligned imagery via a camera recording through an \ac{HMD} is inherently limited, as it does not accurately reflect the user’s true eye perspective. Consistent with prior work~\cite{sutton2022look,Emsenhuber2023}, we use simulated views rendered from the user’s viewpoint using Unity 2022.3 LTS to visualize algorithmic behavior and alignment artifacts in the paper. In the supplemental material, we also include a video of our best effort of capturing \ac{EPR} as seen through the HoloLens~2 display. Furthermore, we conduct our experimental user evaluations, including both qualitative and quantitative measurements, in a real-world scenario using a physical, untethered \ac{OST} \ac{HMD}, allowing us to assess \ac{EPR} performance and accuracy under realistic conditions. 
 
\section{Related Work}
In the following, we provide related work on previous solutions to \ac{EPR}. Subsequently, we discuss use cases that benefit from knowledge of the correct eye-perspective of users through an \ac{OST} \ac{HMD}.

\subsection{User- and Eye Perspective Rendering in AR}
Perspective-correct rendering for \ac{AR} is an issue that is not unique to \ac{HMD}s. It was brought up first in the context of handheld \ac{AR} interfaces~\cite{7893336} such as mobile phones or tablets. The \ac{AR} view of handheld devices uses the integrated device camera view for augmentations, where the viewpoint of the user does not match the camera viewpoint requiring the users to mentally map between the different views. To solve this issue, user-perspective rendering~\cite{Baricevic2012,Baricevic2014} has been proposed that synthesizes a view of the scene based on the user's eye position, transforming the \ac{VST} mobile device into a transparent window onto the scene.

\ac{VST} \ac{HMD}s are also affected by the offset between camera position that captures the view of the scene, and the user's eye position. While nowadays most \ac{VST} \ac{HMD} place the cameras roughly in front of the user's eye, even this small positional offset combined with the individual differences in eye positions are sufficient to lead to a noticeable difference \cite{10.1145/3610548.3618134} and negatively impact depth perception\cite{bailenson2024seeing}. To overcome the positional offset, several approaches have been proposed \cite{10.1145/3384540, 10.1145/3528233.3530701, 10.1145/3588432.3591534} that either compute a 3D reconstruction~\cite{10.1145/3384540} relying on video encoders integrated into the \ac{HMD} to track features between the frames, more recent approaches rely on view generation techniques such as neural rendering \cite{10.1145/3528233.3530701} or light-field rendering \cite{10.1145/3588432.3591534}. Unfortunately, neural and light-field rendering techniques require the computational power of a stationary computer (e.g. a NVIDIA Titan RTX for each eye \cite{10.1145/3528233.3530701}), and, thus, are not applicable to mobile, untethered \ac{HMD} hardware. 

To capture the user's view through an \ac{OST} \ac{HMD}, i.e., the light rays right in front of the user's eyes, prior work explored hardware modifications, such as putting a large sensor with lenslets for capturing incoming light rays right in front of the user's eyes~\cite{10.1145/3588432.3591534}. However, implementing such an approach for \ac{OST} \acp{HMD} would block the actual view of the user. Instead, Langlotz et al.~\cite{Langlotz2016} proposed to place beam splitters in the optical path of the human eye, where they reflect a fraction of the light towards the camera to precisely change the appearance of the environment e.g. to address color-vision deficiency \cite{Sutton2022} or real-world saliency modulation \cite{sutton2022look}. 

Unfortunately, beamsplitters require careful calibration but, more importantly, increase the \ac{HMD}'s size and require extensive hardware modification. To address these shortcomings, recent work started to explore \ac{EPR} for \ac{OST} \acp{HMD}, a software-based method that does not require hardware modifications, and demonstrated the effectiveness of the precise knowledge of the user's view through the displays for view management such as label placement~\cite{Emsenhuber2022,Emsenhuber2023}. We are exploring and extending previous approaches to software-based \ac{EPR} for supporting mobile, untethered \ac{OST} \ac{HMD}s. 

\subsection{Applications of Eye-perspective Rendering}
\ac{AR} applications embed virtual information into the user's view. While applications typically rely on 3D pose information for placing the content (e.g., application windows), more advanced placement algorithms perform image-based analysis on the device's camera feed~\cite{Cao2023} for scene analysis, to adapt content placement to avoid visual interference~\cite{Emsenhuber2022, Emsenhuber2023} or object occlusions~\cite{Jia2021}. For example, recent algorithms using machine learning techniques for AI object detection \cite{Liu2019,Lysakowski2023Yolo,Kim2024,Ghasemi2022DeepLearning} detect and identify individual objects in the scene and provide this information to the user via an \ac{AR} interface~\cite{10.1145/3654777.3676379}. Most of these algorithms work directly on 2D images of the scene, which usually come from a camera attached to the \ac{HMD}. However, mapping the extracted information into the user's view through an \ac{OST} \ac{HMD} leads to inaccuracies and potential ambiguity due to the difference between the camera perspective and the user perspective (\cref{fig:usecases}(Top)(Left)). 

Augmented Vision is another growing area in \ac{AR},  aiming to support individuals with visual impairments or working in low vision environments (e.g., dust, smoke). It commonly relies on 2D imagery to identify scene features that are modulated and enhanced in the \ac{AR} view \cite{Langlotz2024}. For instance, vision support systems for individuals with \ac{CVD} identify critical colors and modulate them via overlays on the \ac{OST} display --- changing color hue and brightness, using patterns that help distinguish them from otherwise similar colors \cite{Sri2011,Sutton2022}, or framing critically colored objects with bounding boxes and additional textual description~\cite{Zhu2023}. Low vision support systems extract edges from camera images using image-based edge detection to emphasize structures in the user's view~\cite{Hwang2014GoogleGlass,Langlotz2024, Htike2021}. However, when naively overlaying in the user's view, the difference in camera perspective and user's eye perspective leads to inaccuracies as the supporting augmentations do not match the real-world scene (\cref{fig:teaser}(Left), \cref{fig:usecases}(Top)(Middle, Right)). 

Aside from practical implications due to the mismatch between camera and user view impacting the usability of the described use cases, the misalignment of \ac{AR} overlays with the real-world scene can cause physiological issues such as nausea and cybersickness ~\cite{chemaly2025mind}. Recent studies suggest that a 15\,mm offset between the camera and the eye is the threshold at which offsets are becoming noticeable~\cite{10.1145/3610548.3618134}. Even with modern, miniaturized cameras and display optics, achieving a camera placement this close to the human eye is hardly feasible, potentially leading to noticeable negative effects when showing information computed in the camera view.

\section{System Overview}
In the following, we present implementation details on three \ac{EPR} techniques included in our open source framework, as well as the image-based analysis approach for our experiment, before illustrating the \ac{EPR} techniques for selected use cases. 

\subsection{Eye-Perspective Rendering}
\begin{figure*}[!tb]
\centering
\includegraphics[width=\textwidth]{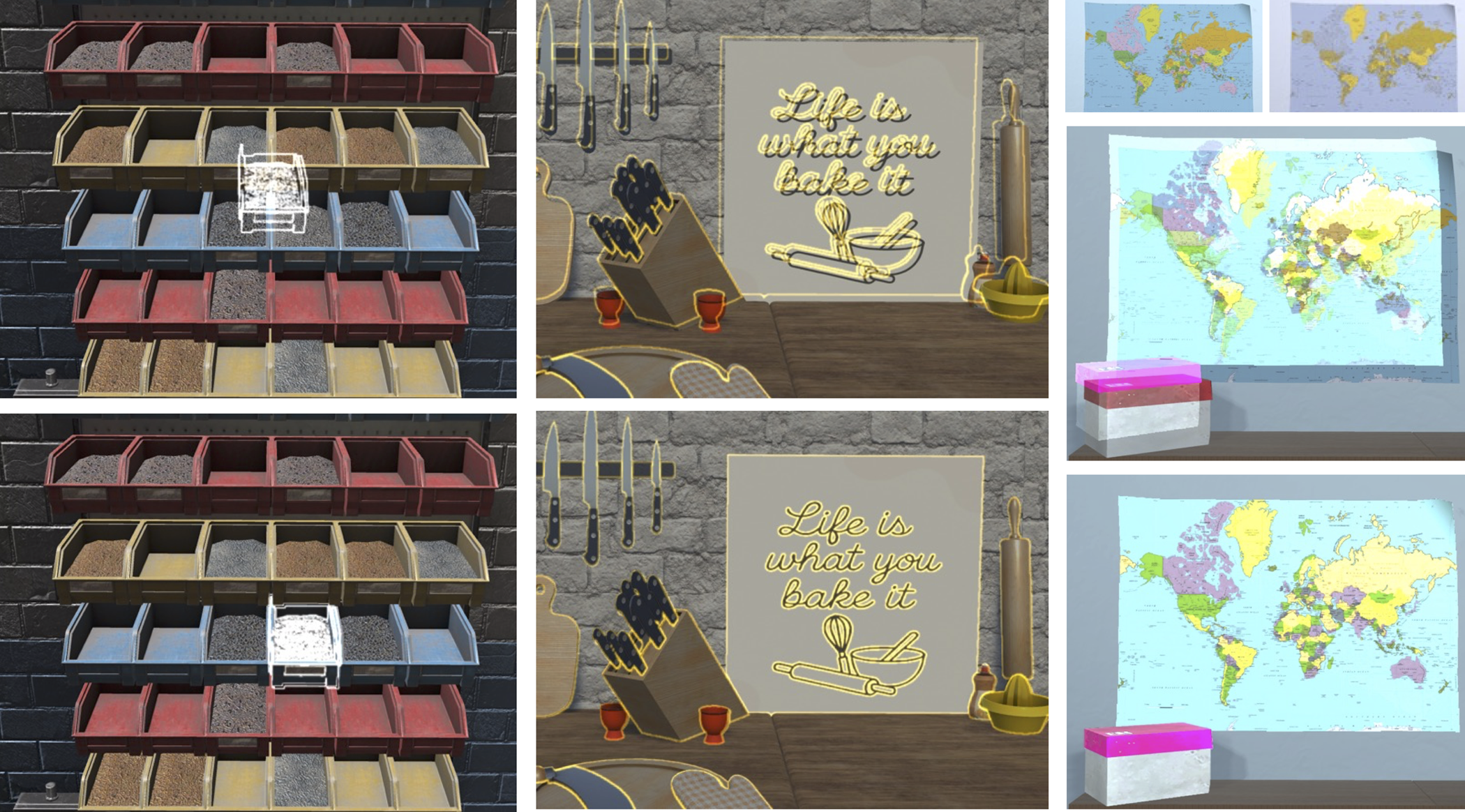}
\caption{Use Cases of Image-based Analysis. (Top) The \eprPlane{} method frequently results in misalignments between overlays and the real-world scene due to the fixed distance of the projection plane. (Bottom) The \eprMesh{} method accurately aligns extracted image information with the real-world scene. Here, we demonstrate the effect for (Left) highlighting outlines of the next picking tray for a maintenance task, (Middle) enhancing scene visibility and text legibility through outline overlays for low vision support, and (Right) applying Daltonization to shift color spaces for individuals with \ac{CVD}. The top images show the original colors and the colors perceived by an individual with \ac{CVD}.}
\label{fig:usecases}
\end{figure*}

Our \ac{EPR} methods apply a texture projection technique to re-render the world-camera view from the user’s perspective using 3D proxy geometry~\cite{Emsenhuber2022, Emsenhuber2023}. Each point on the proxy geometry is first projected into the world-camera view to retrieve the image data available to the \ac{AR} system. The same point is then reprojected into the user's left and right eye views as \ac{EPR}. When the proxy geometry accurately corresponds to the real-world scene geometry, both the world camera and the user perceive consistent visual information. This alignment is illustrated in \cref{fig:teaser}(Right), where a mesh closely matches the scene geometry, and in \cref{fig:teaser}(Middle), where a planar proxy is aligned with the user's gaze target. Correct alignment of \ac{EPR} views for both eyes is essential for achieving proper stereo perception of image-based augmentations, such as object highlighting (\cref{fig:usecases}(Left)). Misalignment between augmentations and the physical scene can introduce perceptual conflicts, potentially degrading the user experience~\cite{Htike2021,Zhu2023}.

Our system is the first to implement multiple \ac{EPR} approaches natively on a state-of-the-art \ac{OST} \ac{HMD}. We use the HoloLens~2 with the Research Mode API to access sensor streams, and Unity with the \ac{MRTK} 2.8.3 for rendering. While developed for HoloLens~2, the \ac{EPR} techniques are generalizable to other platforms, including Magic Leap~2 and Snap AR glasses.

We implement three methods that share the same projection technique but differ in their choice of proxy geometry: a fixed plane for \eprPlane{}, a gaze-aligned plane for \eprGaze{}, and a mesh approximating the scene geometry for \eprMesh{}.

\textbf{\eprPlane.}
In this method, the world-camera view is projected onto a virtual 3D proxy plane placed at a fixed distance in front of the user, parallel to the image plane of the dominant eye. The distance can be adjusted manually, to match a typical working distance. However, due to continuous head motion in mobile use, visual augmentations based on the \ac{EPR} view quickly become misaligned with the real-world scene (\cref{fig:teaser}(Left), \cref{fig:usecases}(Top)), negatively affecting the user experience. While this method is prone to misalignment, it may still be viable for viewing distant scene geometry, where the parallax between the camera and the user’s eyes becomes negligible~\cite{Borsoi2018}. \eprPlane{} is the simplest of the three approaches, requiring no pose tracking, eye tracking, or scene information, and introduces no computational overhead.

\textbf{\eprMesh.}
Instead of a planar proxy, this method uses a spatial mesh reconstruction provided by the \ac{HMD} as a 3D proxy geometry, as proposed by Emsenhuber et al.~\cite{Emsenhuber2023}. The \ac{EPR} view is generated by projecting the world-camera image onto the mesh and re-rendering it from the user's eye position. Since the proxy geometry closely matches the physical scene, \eprMesh{} enables more accurate \ac{EPR} views in which image-based augmentations better align with the real world (\cref{fig:teaser}(Right) and \cref{fig:usecases}(Bottom)). 

The method's effectiveness depends on the quality of the mesh. As discussed by Weinmann et al.\cite{Weinmann2021}, the HoloLens~2’s spatial mesh adequately captures basic indoor architecture, but lacks the precision needed to represent finer scene details, potentially leading to visible mismatches between augmentations and physical objects. The HoloLens~2's spatial mapping accuracy can vary depending on environmental conditions and calibration. Under optimal conditions and with proper ground control, studies have demonstrated that the device can achieve spatial mapping accuracy of 1-2\,cm \cite{Teruggi22HoloLens}. The mesh available through Unity is less precise due to system-level performance constraints. The mesh resolution is governed by the maximum triangles per cubic meter parameter, with a default value of about 500, which is insufficient for capturing finer scene features. In our implementation, we set the parameter controlling this value to "Fine", which corresponds to 2000-3000 triangles per cubic meter, resulting in a runtime rendering performance of approximately 28 fps for stereo \ac{EPR} views.

Alternatively, the \eprMesh{} method could employ depth maps generated via image-based depth estimation~\cite{Rajapaksha2024} or a built-in depth sensor. However, current mobile hardware may not support real-time image-based depth estimation at sufficient quality, and depth sensors often produce incomplete data with missing regions that require computationally intensive inpainting~\cite{Gsaxner2024}. Future \acp{HMD} may implement such features, when the issue of \ac{EPR} becomes more recognized for \ac{OST}.

\textbf{\eprGaze. }
This approach builds on the same plane-projection technique as \eprPlane{}, but dynamically positions the projection plane at the 3D point in the scene the user is currently focusing on. In principle, this point can be determined by computing the eye vergence using a built-in eye tracker. However, since the HoloLens~2 only provides a single gaze ray~\cite{Wang22}, and prior work has shown that vergence estimation degrades with increasing viewing distance~\cite{Mardanbegi2019}, we estimate the gaze target by intersecting the eye-tracked gaze ray with the spatial mesh used in \eprMesh{}. This enables us to position the projection plane at the user's current focus and to generate \ac{EPR} views that are accurate within the user's focal area (\cref{fig:teaser}(Middle)). Misalignments outside this region are less likely to impact the user experience, as the visualization updates when the user’s gaze shifts.

Although our method currently relies on the spatial mesh to estimate depth, it demonstrates the feasibility of the \eprGaze{} approach and its potential for lightweight, scene-agnostic rendering, assuming sufficiently accurate vergence tracking in future \acp{HMD}. Given that vergence estimation is more reliable for nearby objects, this approach is particularly suitable for close-range tasks, where \ac{EPR} accuracy is most critical. For distant objects, parallax between the world camera and the user's eyes becomes negligible, reducing the need for dynamic alignment~\cite{Borsoi2018}.

\subsection{Image Processing}
Our framework provides implementations of basic image-based scene analysis routines, which we use to demonstrate the necessity and impact of \ac{EPR}-aware image analysis in \ac{OST} \ac{AR}, both in functional demonstrations in simulated scenes and in our user study using the HoloLens~2. Implementing mobile versions of the image operations enables both quantitative measurements and qualitative feedback on the effects of \ac{EPR} in real-world scenarios. For our mobile implementation, we avoid computationally intensive object recognition models and instead employ lightweight image processing techniques that allow for real-time manipulation and highlighting of image regions. While we do not currently integrate advanced models, doing so is possible either by executing them on-device if hardware permits, or by streaming world-camera images to an external machine for processing, as shown in previous work~\cite{Bahri2019Yolo}.

All image processing is performed directly on the RGB image captured by the world-facing camera. The resulting annotations or visual transformations are then projected onto the proxy geometry to create the \ac{EPR} view. 
Processing the original camera image rather than the reconstructed \ac{EPR} view avoids artifacts introduced by mesh inaccuracies in \eprMesh{}.
Limited mesh resolution can lead to distorted or low-polygon silhouettes, particularly for small or detailed objects, which in turn degrade the quality of downstream image analysis. By operating on the artifact-free camera image, we preserve processing accuracy.

Our framework implements the following image-based operations as basis for our \ac{EPR} use case demonstrations and user study: 
\begin{itemize} 
\item \textbf{Edge Overlays:} Canny edge detection is applied to the scene to simulate visual support systems for low vision~\cite{Htike2021}. 
\item \textbf{Hue-Based Segmentation:} Objects of interest are highlighted by segmenting regions based on hue and either coloring them or outlining their contours. 
\item \textbf{Daltonization:} The input image is adjusted using color transformations to improve perception for users with \acp{CVD}~\cite{Langlotz2018}. 
\end{itemize}

\begin{figure*}[tb]
\centering
\includegraphics[width=\textwidth]{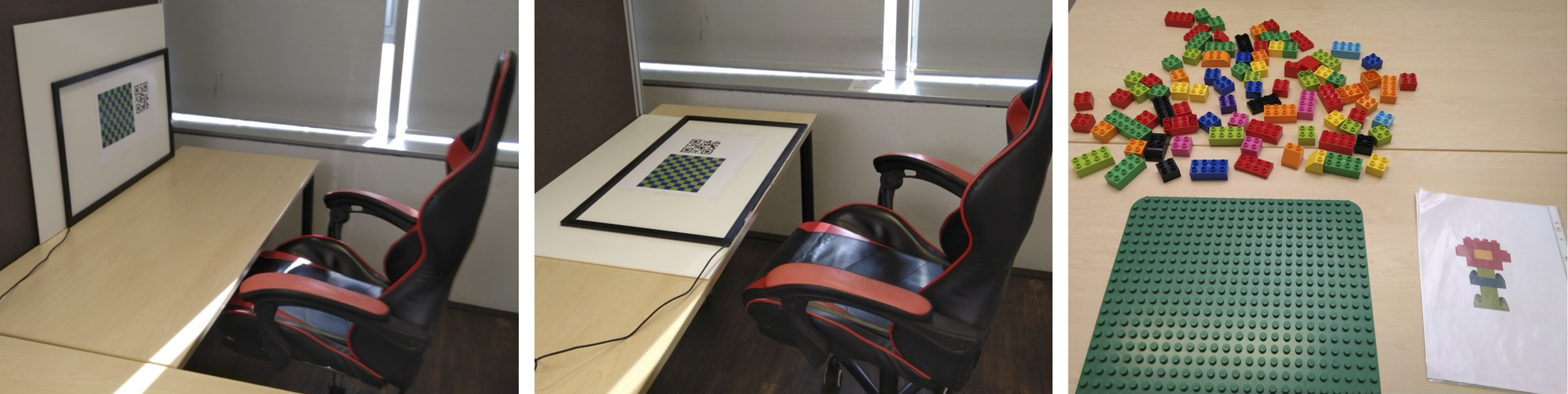}
\caption{Study Setup. (Left) The wall-mounted checkerboard (\condWall), where users were looking at the targets at a $90^\circ$ angle, and (Middle) the checkerboard lying on the table (\condTable), where users were looking at an approximately $50^\circ$ degree angle. (Right) The Duplo playground for the second part of the study.}
\label{fig:study_setup_visualization}
\end{figure*}

\subsection{Use Cases of Image-based Analysis}
In the following, we describe a selection of use cases that benefit from a precise registration of image-based analysis results utilizing an \ac{EPR} method such as \eprMesh{} or \eprGaze{} that aligns results with the viewpoint of the user and the scene.

\textbf{Guidance in Manual Tasks. }
A common use case for \ac{AR} is assisting users in manual tasks, where virtual augmentations indicate the next element to interact with, such as a part~\cite{Eswaran2023ARAssembly}, a tool~\cite{Lai2020}  or a control element of a machine~\cite{Lutnyk2024}. AI-driven, image-based guidance in unstructured scenes is well-suited for such scenarios, where objects of interest are highlighted using their silhouette. 

Highlighting objects with artificial silhouettes, rather than bounding boxes, can convey their shape more clearly and helps distinguish them from similar items. Furthermore, in cluttered environments, bounding boxes may enclose multiple similar objects, causing confusion. \cref{fig:usecases}(Top, Left) illustrates a silhouette highlighting a tray during a sequence of maintenance steps using the naive \eprPlane{} method, where misregistration results in ambiguity. In contrast, \cref{fig:usecases}(Bottom, Left) demonstrates the precise \eprMesh{} approach, where the registered silhouette clearly directs attention to the correct object.

\textbf{Low-vision Support. }
According to the WHO, 2.2 billion people are suffering from a vision impairment~\cite{WHO_Blindness_2023}, which can severely impact the quality of life, particularly by reducing mobility. Edge overlays can improve perception for people with low vision~\cite{Htike2021}, enabling them to detect obstacles more easily, read text, discern image details, and interpret facial expressions. Beyond assisting individuals with low vision, edge overlays can enhance scene perception in environments with limited visibility, such as dusty work settings or smoke-filled areas in firefighter scenarios. 

Simple edge overlays are computationally efficient with image-based analysis. However, previous work has shown that users experience depth perception issues when edge overlays are misaligned with scene geometry and displayed on a fixed-depth plane~\cite{Htike2021} (\cref{fig:usecases}(Top, Middle)). \eprMesh{} allows for precise edge overlays at the correct scene depth, thereby avoiding depth perception conflicts (\cref{fig:usecases}(Bottom, Middle)).

Note that the HoloLens~2 may not be suitable for people having conditions causing low vision due to the fixed focus plane of the display at 2\,m. A retinal projection \ac{HMD}~\footnote{\url{https://retissa.biz/en/}} enables focused augmentations as light is directly focused  onto the retina.

\textbf{Compensating Color Vision Deficiencies. }
\ac{CVD} is a condition in which affected individuals have a reduced ability to distinguish between colors, most commonly between red and green. A common approach to \ac{CVD} compensation is Daltonization, which adjusts colors to reduce combinations that are difficult to differentiate. \ac{OST} \acp{HMD} offer an ideal assistive solution for individuals with \ac{CVD}, allowing a direct modification of the user's view.

Daltonization can be implemented by segmenting problematic colors in the world-facing camera image and overlaying a mask that shifts scene colors to improve color differentiation~\cite{Sri2011, Langlotz2018}. Alternatively, Zhu et al.~\cite{Zhu2023} analyze the camera feed for problematic colors and annotate objects with textual labels and tightly fitting bounding boxes. However, their system renders the output on a fixed-depth plane (1\,m), resulting in depth perception issues and misaligned bounding boxes. Using the same system for Daltonization leads to the same issues (\cref{fig:usecases}(Top)(Right)).

Both Daltonization and bounding geometry benefit from precise \ac{EPR} methods such as \eprMesh{} (\cref{fig:usecases}(Bottom)(Right)), which allow masks or overlays to align with real-world scene geometry at the correct depth, avoiding perceptual inconsistencies.
 
\section{Experiment}
The institutional ethics committee of Salzburg University of Applied Sciences approved this study. We conducted a mixed-method, within-subject user study to compare the three \ac{EPR} approaches in a real-world setting using an untethered, mobile \ac{OST} \ac{HMD} (HoloLens~2). The first part of the study focused on quantifiable performance metrics, requiring participants to interact with the scene based on image analysis from the world camera. In the second part, we gathered qualitative feedback as participants walked through a scene simulating a low vision support system with overlaid edge enhancements. 

\subsection{Quantitative Evaluation}
Participants were asked to identify and touch the correct square on a checkerboard, guided solely by a virtual highlight derived from world camera analysis. The study included two independent variables: \textbf{Orientation} and \textbf{EPR Method}. \textbf{Orientation} had two conditions: \condWall, with the checkerboard mounted on a wall and viewed head-on, and \condTable, with the checkerboard placed flat on a table and viewed at an angle. These conditions allowed us to examine the impact of viewing angle on \ac{EPR} accuracy. \textbf{EPR Method} had three conditions representing our \ac{EPR} implementations: \condFixed, \condGaze{}, and \condMesh. \condFixed{} applied a 20\,cm offset from the scene geometry, simulating a misaligned proxy plane that has been calibrated for a different workspace. Participants were allowed to compensate for this offset by moving their heads.

\textbf{Participants. } We recruited 18 participants (8 female, $\overline{X}$$=$33 (8) years). On a scale from one (worst) to seven (best), the mean of self-rated \ac{AR} experience was 2.5. All participants had normal or corrected-to-normal vision.

\begin{table*}[!ht]
   \scriptsize
     \begin{tabularx}{\textwidth}{l||X|X|X|X}
         & EPR Method & \condMesh{} vs \condFixed{} & \condMesh{} vs \condGaze{} & \condFixed{} vs \condGaze{}  \\            
      \hline
      \hline
      Correctness & \art{2}{85}{129.55}{\textless{}.001} & \artcon{84}{14.49}{\textless{}.001}  & -  & \artcon{84}{13.32}{\textless{}.001} \\
      Task Completion Time & \art{2}{85}{41.9}{\textless{}.001} & \artcon{85}{8.79}{\textless{}.001} & \artcon{84}{2.19}{=0.032} & \artcon{85}{6.6}{\textless{}.001}  \\
      \hline
      \hline 
      SEQ & \art{2}{85}{33.7}{\textless{}.001} & \artcon{84}{7.58}{\textless{}.001} & - & \artcon{85}{6.52}{\textless{}.001} \\
      TLX Overall & \art{2}{85}{61.4}{\textless{}.001} & \artcon{85}{9.97}{\textless{}.001} & - & \artcon{85}{9.17}{\textless{}.001} \\
      Mental Demand & \art{2}{85}{58.3}{\textless{}.001} & \artcon{84}{9.55}{\textless{}.001}& - & \artcon{84}{9.13}{\textless{}.001}  \\
      Physical Demand  & - & - & - & - \\
      Temporal Demand & \art{2}{85}{3.8}{=0.026} & \artcon{85}{2.72}{=0.024} & - & -  \\
      Performance & \art{2}{85}{83.7}{\textless{}.001} & \artcon{85}{10.71}{\textless{}.001}& - & \artcon{85}{11.64}{\textless{}.001} \\
      Effort & \art{2}{85}{35.1}{\textless{}.001} & \artcon{84}{7.56}{\textless{}.001}& -& \artcon{85}{6.9}{\textless{}.001}  \\
      Frustration & \art{2}{85}{14.01}{\textless{}.001} & \artcon{85}{4.94}{\textless{}.001} &  - & \artcon{85}{4.13}{\textless{}.001}  \\
    \end{tabularx}    
    \caption{Statistically significant results of ART and ART contrasts for main effects EPR method. }
    \label{tab:art_con_results_overall_filtered}
\end{table*}

\textbf{Task. } 
During each trial, participants interacted with a checkerboard consisting of 11$\times$9 squares, each measuring 2\,cm. An algorithm randomly selected a square, highlighting it with a white outline. Participants were instructed to identify and touch the center of the matching real square, then return their hand to a starting position on their legs. The task was repeated 30 times per \ac{EPR} method, with the first 5 repetitions discarded to reduce learning effects.

\textbf{Apparatus. } 
All experiments were conducted with participants seated on an adjustable chair to ensure a consistent viewing angle across different body heights. Participants adjusted their chair to comfortably align their eyes with the middle row of the checkerboard in \condWall, positioned at a height of 103\,cm (\cref{fig:study_setup_visualization}(Left)). For \condTable, the chair height remained unchanged, with the table set at 75\,cm, resulting in an approximate viewing angle of $50^\circ$ (\cref{fig:study_setup_visualization}(Middle)). The checkerboard was centered relative to the participant’s dominant hand.
To support square detection and highlighting, we developed a custom \ac{AR} application using Unity, the \ac{MRTK}, and our framework’s image analysis functions. For \eprMesh{} and \eprGaze{}, we used the built-in HoloLens~2 eye calibration tool to obtain accurate eye position data necessary for computing \ac{EPR} views. 
Touch interactions were recorded using a 70.6$\times$39.7\,cm infrared touch frame (GreenTouch) with 2\,mm precision, connected to a PC. A PC-side application logged touch event timestamps and communicated with the HoloLens~2 via TCP. Upon receiving a touch event, the \ac{AR} app selected and highlighted a new square, pausing for 3~seconds to allow the participant to return their hand to the resting position.
Participants were free to move their upper bodies and heads, which occasionally caused the highlighted square to leave the HoloLens~2’s field of view.
To determine whether the checkerboard remained within the participant’s view, we placed a QR code next to the grid and tracked head position and orientation. When the grid was not visible, the application paused and displayed directional arrows to guide participants back to the correct viewing position.
The study was conducted in a room with controlled lighting (LUPO Superpanel Dual Color~60) to ensure consistent conditions across sessions. Ambient light was measured at 317~lux at a color temperature of 3220 Kelvin using a Mavospec Base spectrometer.
  
\textbf{Data Collection. }
We measured \ac{TCT} as the time between highlighting a new square on the HoloLens~2 and the participant touching the corresponding square, as detected by the touch frame. Accuracy was defined as the percentage of correctly touched squares. Task load was assessed using the NASA TLX~\cite{HART1988}, and usability via the \ac{SEQ}. Participants also ranked the \ac{EPR} methods by preference.
To estimate alignment accuracy between the \ac{EPR} view and the real environment, participants viewed a second checkerboard mounted on a wall, with 1\,cm squares overlaid using Canny edge detection and white silhouettes. Positioned 75\,cm from the wall, participants estimated the offset between the real and virtual squares in millimeters for \eprMesh{} and \eprGaze{}. \eprPlane{} was excluded, as it's accuracy relies on continuous manual alignment.

\textbf{Procedure. }  After welcoming participants, they signed an informed consent form and filled out a demographic questionnaire. Their dominant eye was determined using the Miles Test~\cite{Miles1930} to align the proxy planes of \condGaze{} and \condFixed{} accordingly. Participants then put on the HoloLens~2 and used the built-in eye calibration tool to adjust the device to their eyes.

Participants were seated in the task area and instructed to touch the highlighted square as quickly as possible once it became visible on the \ac{HMD}, and to move the hand back into its resting position afterwards. Orientation and \ac{EPR} method were counterbalanced. Half of participants started with \condTable, the other half with \condWall. \textbf{EPR Method} was balanced using a 3x3 Latin Square Table. After finishing 30 repetitions of an \ac{EPR} method, participants filled out NASA TLX and SEQ questionnaires, before continuing with the next \ac{EPR} method. After finishing all \ac{EPR} conditions, participants ranked the \ac{EPR} methods and continued with the next Orientation condition. 

Participants were allowed to remove the \ac{HMD} when filling out the questionnaires, or to rest. If the \ac{HMD} was removed, the eye calibration was performed before the next condition started.  After finishing all conditions, participants were asked to estimate the accuracy of the alignment of \eprMesh{} and \eprGaze{} using another wall-mounted checkerboard. Participants then continued with the second, qualitative part of the study. We collected 2 (Orientation) x 3 (\ac{EPR} method) x 25 = 150 data points for each participant, leading to 2700 repetitions over all 18 participants. 

\textbf{Hypotheses. }
\begin{itemize}
\item \textbf{H1.} Due to the alignment of the visualization with the correct depth at the focus point of the user, \condMesh{} and \condGaze{} outperform \condFixed{} in \textbf{\condWall}. We did not expect differences between \condMesh{} and \condGaze{} due to correct alignment.
\item \textbf{H2.} Due to the alignment of the visualization with the correct depth at the focus point of the user, \condMesh{} and \condGaze{} outperform \condFixed{} in \textbf{\condTable}. We did not expect differences between \condMesh{} and \condGaze{} due to correct alignment.
\item \textbf{H3.} \condFixed{} performs better for  \condWall{} than for \condTable{} as the foreshortening effect of the steeper viewing angle further aggravates the misalignment.
\item \textbf{H4.} Participant prefer \condMesh{} and \condGaze{} over \condFixed, as aligning virtual content to the correct depth reduces perceptual depth conflicts.
\end{itemize}

\textbf{Results. } 
We used the statistics software \emph{R}, data was evaluated with a significance level of 0.05. The residuals did not fulfill the normality requirement. Therefore, we utilized align-and-rank transform (ART)~\cite{Wobbrock2011} and follow-up ART contrasts~\cite{Elkin2021} for post-hoc analysis. The reported p-values are Bonferroni-Holm corrected. 
For each \ac{EPR} method and Orientation condition, we calculated the mean over all task conditions for each participant. Descriptive statistics are summarized as box plots in 
\cref{fig:boxplots}, as well as tables in supplemental material. Ranking results are shown in \cref{fig:ranking}.
Statistically significant differences between \ac{EPR} methods and Orientation conditions are presented in Table~\ref{tab:art_con_results_overall_filtered}. 
In the following, we report on interaction effects between \ac{EPR} method and Orientation. ART revealed significant differences for correctness ~\arttext{2}{85}{19.2}{$<$.001}, contrasts between \condWall{}*\condFixed{} and 
\condWall{}*\condMesh{}~\artcontext{85}{8.7}{$<$.0001}{}, 
\condWall{}*\condGaze{}~\artcontext{85}{7.7}{$<$.0001}{}, 
\condTable{}*\condMesh{}~\artcontext{85}{9.4}{$<$.0001}{}, and 
\condTable{}*\condGaze{}~\artcontext{85}{9.6}{$<$.0001}{}, as well as
between 
\condTable{}*\condFixed{} and 
\condWall{}*\condMesh{}~\artcontext{85}{11.2}{$<$.0001}{}, 
\condWall{}*\condGaze{}~\artcontext{85}{7.7}{$<$.0001}{}, 
\condTable{}*\condMesh{}~\artcontext{85}{11.9}{$<$.0001}{}, and 
\condTable{}*~\condGaze{}~\artcontext{85}{12.1}{$<$.0001}{}.
ART revealed significant differences for \ac{TCT} ~\arttext{2}{85}{5.7}{$=$.005}, contrasts between 
\condWall{}*~\condFixed{} and 
\condWall{}*~\condMesh{}~\artcontext{85}{8.0}{$<$.0001}{},
\condWall{}*~\condGaze{}~\artcontext{85}{7.9}{$<$.0001}{}, 
\condTable{}*~\condMesh{}~\artcontext{85}{6.7}{$<$.0001}{}, and 
\condTable{}*~\condGaze{}~\artcontext{85}{3.8}{$=$.002}{}, as well as between 
\condTable{}*~\condFixed{} and  
\condWall{}*~\condMesh{}~\artcontext{85}{6.1}{$<$.0001}{}, 
\condWall{}*~\condGaze{}~\artcontext{85}{5.97}{$<$.0001}{}, and
\condTable{}*~\condMesh{}~\artcontext{85}{4.8}{$=$.0001}{}. Furthermore, between 
~\condTable{}*~\condGaze{} and 
~\condWall{}*~\condGaze{}~\artcontext{85}{4.05}{$=$.0009}{}, 
~\condWall{}*~\condMesh{}~\artcontext{85}{4.2}{$=$.0006}{}, and 
~\condTable{}*~\condMesh{}~\artcontext{85}{2.9}{$=$.027}{}.
In terms of \textbf{alignment accuracy}, for \eprMesh, participants determined a mean offset of 1.1\,mm (sd$=$0.65\,mm), for \eprGaze{}, a mean offset of 1.3\,mm (sd$=$0.7\,mm). 

\begin{figure*}[tb]
\centering
\includegraphics[height=10cm]{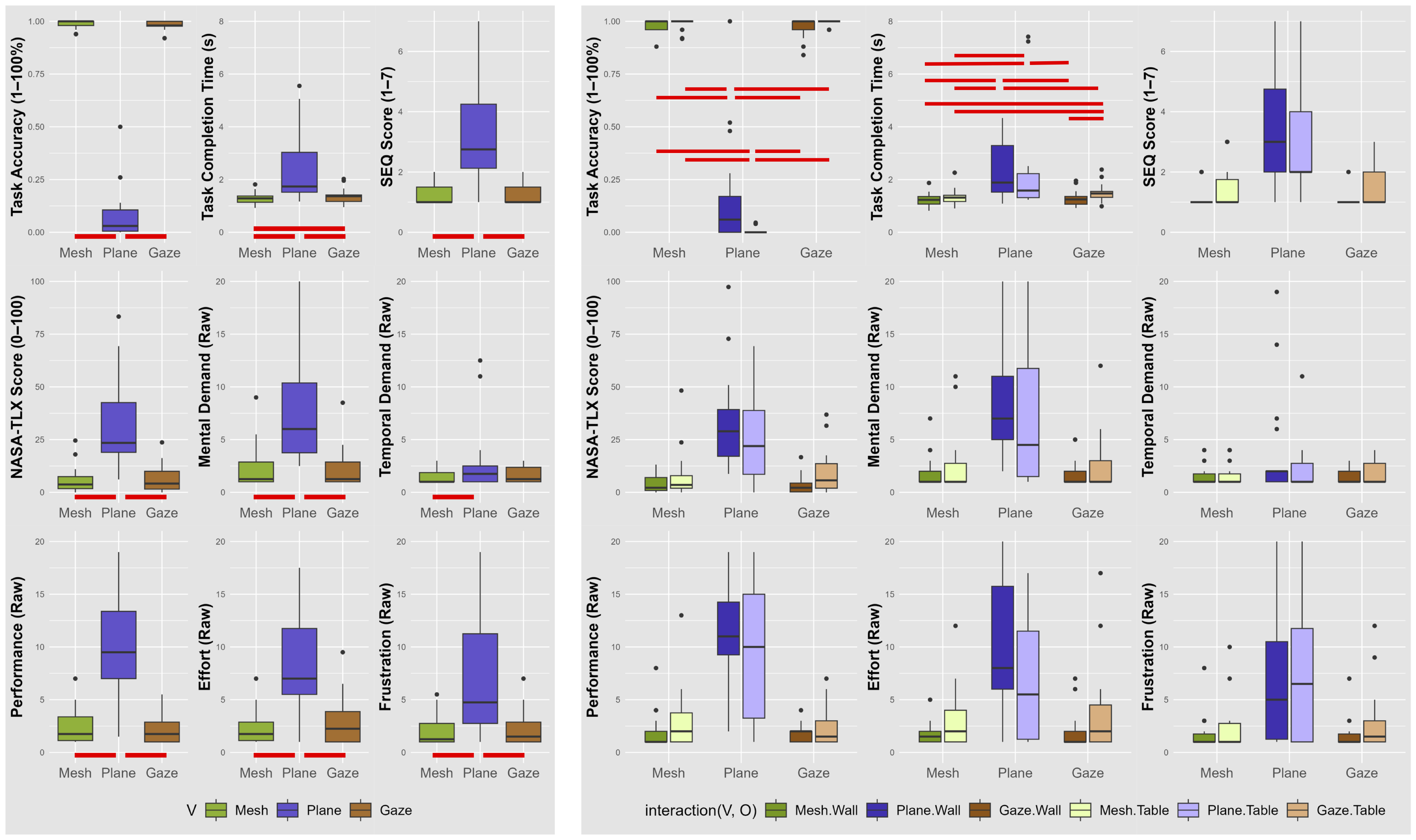}
\caption{Box Plots of Overall EPR Methods Conditions and Orientation * EPR Method Conditions. Significant differences are indicated with horizontal lines. 
}
\label{fig:boxplots}
\end{figure*}

\begin{figure}[tb]
\centering
\includegraphics[width=\columnwidth]{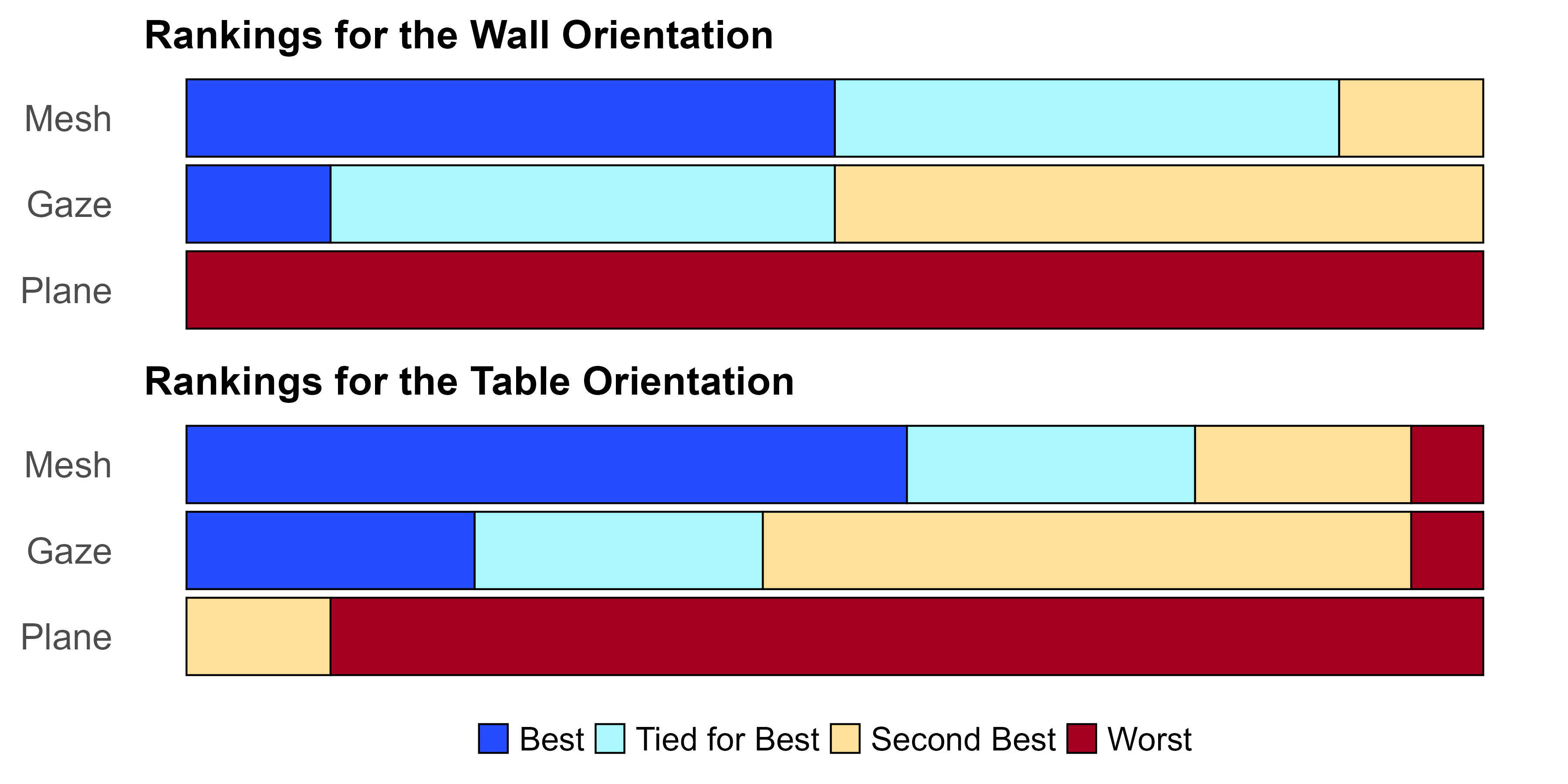}
\caption{Ranking of EPR Methods for both Orientations. 
}
\label{fig:ranking}
\end{figure}
 
\subsection{Qualitative Feedback}
In the second part of the study, we collected qualitative feedback in a use case corresponding to a low vision support system where edges are highlighted to emphasize structures in the environment.

\textbf{Apparatus. }
We calculated edge overlays on the HoloLens~2 by applying Canny edge detection to the world-camera video feed. The edges were shown in white color in the field of view of the user using the three \ac{EPR} methods (\condFixed, \condGaze, \condMesh). Similar applications have been proposed, to help people with low vision \cite{Htike2021, Hwang2014GoogleGlass}. However, previous work projected edge enhanced image onto a proxy plane at a fixed distance in front of the user, which corresponds to the \condFixed{} condition. Expecting misalignment between the overlaid edges and the scene, we allowed users to manually adjust the distance of the proxy plane via a menu. Participants had the option to switch between all three \ac{EPR} methods at any time.

\textbf{Procedure. }
Participants were encouraged to walk around the study room while trying out the different \ac{EPR} methods and conveying their impressions of them. We then asked them to read a Snellen chart on a wall while trying out the three methods. The aim of this was to see if the \ac{EPR} methods were robust enough the be used for enhancing the reading process by highlighting certain letters or key words, e.g. as proposed by \cite{Toyama2013}. Finally, we asked them to sit down at a table where Duplo bricks were laid out (\cref{fig:study_setup_visualization}(Right)). The participants were then asked to build three simple shapes out of the available Duplo bricks, one for each \ac{EPR} method. The shapes consisted of 10 bricks each. Participants used printed images of the shapes as guidance (see supplemental material).  

\textbf{Data Collection. } We collected observations and user feedback during this part of the study, while participants worked through the different tasks in the scene. Participants were also encouraged to think-aloud while trying out the \ac{EPR} methods. 

\textbf{Results. } In the following, we present participant feedback and observations, clustered by common themes that emerged. 

\textbf{Use Case: Low Vision. } Participants approached the low vision support system in different ways based on their own experience. There was a difference between participants with and without vision issues (e.g., myopia). Seven of the participants wore glasses and one participant mentioned having undergone corrective surgery in the past. These participants were overall more receptive to the idea of using the edges as a reading aide. Four participants mentioned the support of outlines: \textit{"It helps with the clarity"} (P7), \textit{"The contrast is better"} (P16). Participants also opined usage scenarios: \textit{"I think it would be helpful outdoors, for recognizing street signs"} (P8), \textit{"You could use it for street signs"} (P17), \textit{"Could be useful for people with color vision deficiencies"} (P16).

\textbf{Use Case: Visual Guidance.} 
When participants interacted with the edge overlays of the image-based \ac{EPR} view analysis in the second part of the study, participants mentioned use cases that related to visual search tasks. Use Cases were mostly related to search tasks: \textit{"It could help me find my glasses"} (P16),  \textit{"It could help me find things that I misplaced"} (P17), \textit{"It could be used to point out the right tools"} (P12). Participants pointed out that for a search task more selective highlighting of objects would be beneficial.

\textbf{Selective Highlighting.}  
During the Duplo building task, only two participants (P8, P4) found the outlines helpful in identifying bricks, which is understandable as the visualization is inspired from low vision support and not guidance. Hence, participants mentioned improvements, such as making the outlines customizable (color, thickness, brightness), offering additional modes (shape-filling instead of outlining), reducing the number of visible outlines, or integrating the outlines into a more intelligent application (offering step-by-step instructions). 
 
\textbf{Failure of \eprPlane.}
Participants could freely experiment with all three \ac{EPR} methods. However, all participants eventually focused on either \eprGaze{} or \eprMesh{} on their own accord. Despite explanations regarding the functioning of the \eprPlane{}~\cite{Htike2021, Zhu2023} and that it could be aligned manually by setting the distance via an \ac{AR} button, or by simple head motion, participants were very expressive in pointing out the limitations of the technique: \textit{"The outlines don’t fit at all"} (P2), \textit{“The fixed distance method is just nonsense. It’s too annoying to adjust it all the time”} (P0), \textit{“Fixed is horrible!”} (P7). When reading from the Snellen chart, participants completely disregarded \eprPlane. During the Duplo building task participants further pointed out:  \textit{"It’s making everything worse"} (P11), \textit{"It is a catastrophe"} (P0), \textit{"I can see the plane, it’s blocking my view of the bricks"} (P5), \textit{"It is not helping, it’s actively hindering me"} (P7).

\textbf{Perceptual Issues.} A common criticism was the lack of temporal coherence in edge overlays across all \ac{EPR} methods, during head motion (P1, P8, P16). The simple Canny edge detection also lead to inconsistencies in outlining differently colored Lego bricks making it harder to distinguish between them for seven participants.

\begin{figure}[tb]
\centering
\includegraphics[width=\columnwidth]{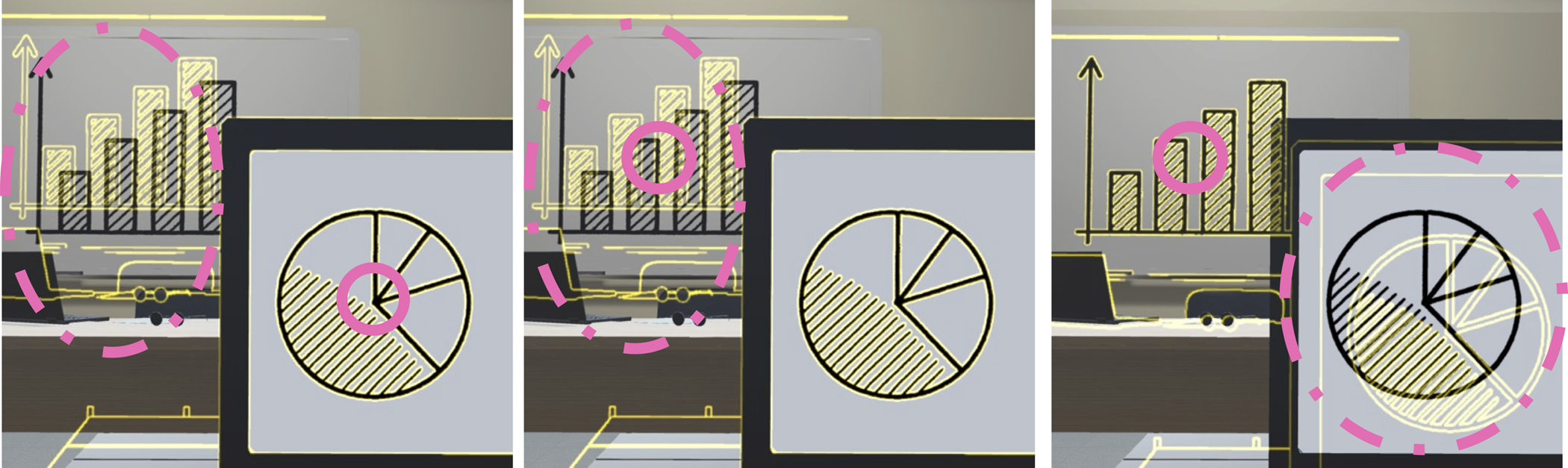}
\caption{Adaptation Process of \eprGaze. Adapting the proxy plane to a new depth using eye tracking potentially introduces temporary ambiguities due to system latency. (Left) As the user focuses on the foreground (purple circle), the proxy plane is aligned with the foreground geometry. The edge highlights in the background are misaligned (dashed lines). (Middle) When the user changes their gaze focus to the background (circle), they may perceive a short misalignment of highlights and background (dashed) before the system detects the user's intention, and (Right) adapts the proxy plane to the new depth. 
}
\label{fig:gazeproxyupdate}
\end{figure}

\section{Discussion}
In the following, we discuss results in relation to our hypotheses, before presenting our insights in more detail.

\subsection{Hypotheses}
The main effects between the \ac{EPR} methods were statistically significant so that overall \condMesh{} and \condGaze{} outperformed \condFixed{} in all measures. However, there were almost no additional interaction effects for ART for (\ac{EPR} method * Orientation) aside from accuracy and \ac{TCT}. 

\textbf{H1.} For \condWall, \textbf{we accept H1}, as \condMesh{} and \condGaze{} clearly outperformed \condFixed{}, which displayed the image-based analysis at a fixed distance from the user, both overall (\cref{fig:boxplots}(Left)) and when analyzing interaction effects in terms of \ac{TCT} and accuracy (\cref{fig:boxplots}(Right)).
 
\textbf{H2.} For \condTable, \textbf{we partially accept H2}, as \condMesh{} and \condGaze{} again outperformed \condFixed{} across all measures (\cref{fig:boxplots}(Left)) and in interaction effects for \ac{TCT} and accuracy (\cref{fig:boxplots}(Right)).

However, \ac{TCT} for \condGaze{}*\condTable{} was significantly different from \condGaze{}*\condWall{}, \condMesh{}*\condTable{}, and \condMesh{}*\condWall{}. We partly attribute this to a perceivable delay when updating the proxy plane depth, due to a combination of eye tracker latency and the larger misalignment between the proxy plane that is orthogonal to the user’s view and the horizontal table surface compared to \condWall{}, which potentially requires users to reorient as the visualization adapts (\cref{fig:gazeproxyupdate}).

\textbf{H3.} Although \condFixed{} performed better on average in the \condWall{} condition in terms of accuracy, \textbf{we reject H3}, as the difference was not statistically significant. Despite participants being allowed to adjust their head position after the task began, only two were able to compensate for the offset caused by the misalignment between the plane and real-world geometry. While we did not observe a significant difference for \condFixed{} across Orientations, we found a significant \ac{TCT} difference between \condGaze{}*\condTable{} and \condGaze{}*\condWall{} as described in the discussion of H2. 

\textbf{H4.} \textbf{We accept H4}, as participants clearly preferred \condMesh{} and \condGaze{} in both Orientation conditions (\cref{fig:ranking}). Some participants could not distinguish between \condMesh{} and \condGaze{} in terms of quality and ranked both equally. This was more common in the \condWall{} condition ($n=7/18$) than in \condTable{} ($n=4/18$), which may indicate that temporary misalignments in \ac{EPR} results with \condGaze{} due to the plane adaptation process(\cref{fig:gazeproxyupdate}) were more noticeable in \condTable{} (see discussion of H2 and H3).

\subsection{Insights}
Based on the quantitative data, qualitative feedback, and observations, we present the following insights.

\textbf{\eprGaze{} as Effective Alternative.} Overall, \eprGaze{} performed comparably to \eprMesh{}, which depends on detailed mesh reconstruction being available. Participants estimated only minor alignment errors of approximately 1\,mm with the real world in both cases. Thus, \eprGaze{} represents a viable alternative to \eprMesh{}, requiring no additional scene information such as depth maps or mesh reconstructions when relying on precise eye tracking for vergence calculation~\cite{Wang22,Mardanbegi2019}.
Since the aim of \eprGaze{} is to create localized alignment directly in the area around the user's focus point by approximating scene geometry in that area with a plane, scene depth variations will cause misalignments between the virtual content and the real world. To avoid this, the gaze target should fill out the area in the user's field of view with a uniform depth. In our current implementation we did not enforce a specific size for the gaze target, however participants did not mention noticing such alignment errors during the study, e.g. when interacting with the Duplo bricks. However, future work should determine the minimum size at which the gaze target must fill out the user’s visual field to ensure perceptually seamless alignment, potentially on the order of the human foveal region (approximately 1–2° of visual angle) in line with foveated rendering~\cite{Albert2017Foveated}.

\textbf{\eprGaze{} using Vergence-based Depth Estimates.}
Although our results show that \eprGaze{} is viable, our implementation relies on the spatial mesh to estimate gaze-target distance because the HoloLens~2 provides only a single combined gaze ray, preventing direct vergence calculations~\cite{Wang22}. Placing the proxy plane via the intersection of the spatial mesh and the gaze ray can introduce errors e.g. when the spatial mesh does not reflect changes in the environment fast enough. Using an eye tracker that directly measures vergence distance avoids these issues as \eprGaze{} does not rely on a potentially imprecise or slowly updating mesh. Vergence-based depth estimates are more precise at smaller distances and degrade with distance due to angular errors~\cite{Mardanbegi2019} making \eprGaze{} particularly suitable for close-range tasks. For distant scene geometry, the parallax between the world camera and the user's eyes becomes minimal~\cite{Borsoi2018}, and continuous \ac{EPR} updates may not be required at ranges where vergence-based estimates are less precise. Further experiments are needed to fully explore the potential of vergence-based \eprGaze{} at different distances.

\textbf{Delayed Eye Tracker Updates.} Our experiments show reduced \ac{TCT} in \condGaze{}*\condTable{} compared to \condGaze{}*\condWall{} and both \condMesh{} conditions. This can be explained by the HoloLens~2’s eye-tracking refresh rate (30 Hz) combined with the \ac{MRTK} gaze-pointer implementation requiring at least 40 gaze samples for stabilization. These factors introduce a noticeable delay whenever the proxy plane is realigned via gaze refocus (\cref{fig:gazeproxyupdate}).
In \condGaze{}*\condWall{}, the proxy plane’s angular alignment with the grid meant that gaze shifts toward the highlighted square required minimal depth adjustment, masking the delay. In \condGaze{}*\condTable{}, the poorer alignment from the viewing angle caused larger depth adjustments, forcing participants to wait for the proxy plane to update before touching the highlighted square. Several participants reported noticing this delay with \eprGaze{}.
Replacing the \ac{MRTK} gaze pointer with a custom implementation using fewer samples can improve the delay, but risks of reducing pointer stability due sensor jitter and head movements not being smoothed. Further research is needed to fully understand how gaze-based proxy-plane updates affect task performance and to quantify the improvements achievable with higher-refresh-rate eye trackers.

\textbf{High Accuracy of \ac{EPR}.} Since capturing the user's \ac{EPR} view is not possible with an \ac{OST} \ac{HMD}, we asked participants to estimate the alignment error of edge overlays. While this is a subjective measure, it enhances ecological validity compared to using instrumentation such as a head rest by allowing unrestricted user movement when wearing the \ac{HMD}. At 75\,cm from a wall, participants reported alignment errors of approximately 1\,mm for both \eprMesh{} and \eprGaze{}, indicating that \ac{EPR} supports precise image-based vision augmentation.

Alignment accuracy was evaluated using only a single \ac{HMD} model (HoloLens~2). However, given the low computational overhead of the proposed software-based \ac{EPR} methods, we are confident in their applicability to a broad range of \acp{HMD}. Nonetheless, future investigations should explore \ac{EPR} performance across a more diverse set of devices, including wearables, e.g. Snap glasses.

\textbf{Correct Depth Perception.} Most participants using \eprPlane{} reported seeing double overlays, i.e., one virtual square highlight for each eye. This visual offset arises from the misalignment between the user’s real-world focus point and the fixed proxy plane. As a result, participants could not focus on both simultaneously. Participants did not report double vision for \eprMesh{} and \eprGaze{}, which underlines the importance of aligning image analysis with the correct scene depth.

\textbf{Disregard Fixed-distance \eprPlane{}.} \eprPlane{} performed significantly worse than other techniques, and, like in previous work~\cite{Zhu2023,Htike2021}, participants strongly criticized it. Despite being instructed on how to compensate for the fixed offset, either via a software setting or head movement, participants found \eprPlane{} impractical and ineffective for any given tasks. Therefore, we strongly advise researchers and practitioners to avoid fixed-distance \eprPlane{} for image-based vision augmentation on \ac{OST} \acp{HMD} and use depth-aligned methods such as \eprMesh{} or \eprGaze{} instead.

\textbf{Optimize Object Highlights.} For the qualitative study, we used Canny edge detection with a fixed highlight color to simulate a low vision support system. Participants noted that the edge overlays and chosen color sometimes obscured object recognition. Future research should explore adaptable color schemes that achieve a good contrast to the scene~\cite{Gabbard2007}. Furthermore, due to the precision of \ac{EPR}, future systems can move beyond coarse bounding boxes~\cite{Zhu2023} and use fine-grained silhouettes or minimal augmentations like points or lines rendered within a small object of interest, potentially reducing clutter introduced by additional augmentations. In procedural tasks (e.g., training, assembly), filtering techniques can further reduce visual clutter of highlights.

\textbf{Need for Stable Image Analysis.} Like in prior work~\cite{Htike2021}, participants experienced issues with the temporal stability of edge overlays. Our qualitative feedback scenario used a basic Canny edge implementation, which led to flickering due to constant recomputation. Limited by the computational and battery constraints of the HoloLens~2, we could not deploy more advanced algorithms in this experiment. As mobile hardware improves or shifts to edge/cloud processing, future systems may support more stable, intelligent vision augmentation~\cite{Bahri2019Yolo}.

\section{Conclusion}
In this paper, we compared three \ac{EPR} methods to collect quantitative and qualitative measures for vision augmentation for \ac{OST} \acp{HMD} that need alignment with a real world scene. Our results clearly showed that the commonly used \eprPlane{} approach that places a proxy plane at a fixed distance from the user~\cite{Zhu2023, Htike2021, Hwang2014GoogleGlass} fails due to severe misalignments of the analyzed scene information and the users view. 

Researchers and practitioners should focus on \ac{EPR} methods that align the augmentations with the corresponding real-world geometry within the field of view of the user instead of the world-camera view. \eprMesh{} achieves this via a depth map from an RGBD camera or a mesh proxy based on a scene reconstruction, while our novel \eprGaze{} approach moves the proxy plane to the depth the user currently focuses on. We find that \eprGaze{} provides an effective alternative to the computationally more complex \eprMesh{}, and can be realized without scene geometry given a sufficiently precise eye tracker for eye vergence depth detection~\cite{Matsuda2013}. 
 
We provide our mobile \ac{EPR} methods and study framework as open source to foster future research for vision augmentations for mobile, untethered \ac{OST} \acp{HMD} in real-world environments. \ac{EPR} is highly relevant for scenarios that only rely on image-based analysis~\cite{Cao2023} and utilize advanced AI-based \acp{LLM} and \acp{VLM}~\cite{10.1145/3654777.3676379}, which require that results are rendered from the user's eye perspective. 

While we propose a software-based framework for \ac{EPR} to stimulate further research, long-term, we see the need to integrate \ac{EPR} natively into \ac{OST} \acp{HMD} either via efficient compute hardware specialized for this task, or novel hardware architectures~\cite{Langlotz2016, Langlotz2018}. Our hope is that if more research explores the effectiveness of \ac{EPR} for real-world use cases, solution providers will pick up \ac{EPR} as a standard feature for \ac{OST} \ac{HMD} so that future applications can make full use of advanced image analysis models to support users in their everyday tasks.

\section*{Supplemental Materials}
The open source repository for the \ac{EPR} framework: \url{https://github.com/DigitalRealitiesLab/MobileEPR}

\acknowledgments{
This work was supported by a grant from the Austrian Research Promotion Agency (grant no. 877104).
}

\bibliographystyle{abbrv-doi-hyperref}

\bibliography{template}

\begin{thebibliography}{10}

\bibitem{Albert2017Foveated}
R.~Albert, A.~Patney, D.~Luebke, and J.~Kim.
\newblock Latency requirements for foveated rendering in virtual reality.
\newblock {\em ACM Transactions on Applied Perception (TAP)}, 14(4):1--13, 2017.

\bibitem{Sri2011}
B.~S. Ananto, R.~F. Sari, and R.~Harwahyu.
\newblock Color transformation for color blind compensation on augmented reality system.
\newblock In {\em 2011 International Conference on User Science and Engineering (i-USEr )}, pp. 129--134, 2011. \href{https://doi.org/10.1109/iUSEr.2011.6150551}
{doi: {{%
10\hspace{.1pt}\discretionary{.}{%
}{.}\hspace{.4pt}1109\discretionary{/}{%
}{/}iUSEr\hspace{.1pt}\discretionary{.}{%
}{.}\hspace{.4pt}2011\hspace{.1pt}\discretionary{.}{%
}{.}\hspace{.4pt}6150551}}}


\bibitem{Bahri2019Yolo}
H.~Bahri, D.~Krčmařík, and J.~Kočí.
\newblock Accurate object detection system on hololens using yolo algorithm.
\newblock In {\em 2019 International Conference on Control, Artificial Intelligence, Robotics \& Optimization (ICCAIRO)}, pp. 219--224, 2019. \href{https://doi.org/10.1109/ICCAIRO47923.2019.00042}
{doi: {{%
10\hspace{.1pt}\discretionary{.}{%
}{.}\hspace{.4pt}1109\discretionary{/}{%
}{/}ICCAIRO47923\hspace{.1pt}\discretionary{.}{%
}{.}\hspace{.4pt}2019\hspace{.1pt}\discretionary{.}{%
}{.}\hspace{.4pt}00042}}}


\bibitem{bailenson2024seeing}
J.~N. Bailenson, B.~Beams, J.~Brown, C.~DeVeaux, E.~Han, A.~C.~M. Queiroz, R.~Ratan, M.~Santoso, T.~Srirangarajan, Y.~Tao, and P.~Wang.
\newblock Seeing the {World} {Through} {Digital} {Prisms}: Psychological {Implications} of {Passthrough} {Video} {Usage} in {Mixed} {Reality}.
\newblock {\em Technology, Mind, and Behavior}, 5(2: Summer 2024), 2024.

\bibitem{Baricevic2014}
D.~Bari{\v{c}}evi{\'{c}}, T.~H{\"{o}}llerer, P.~Sen, and M.~Turk.
\newblock {User-perspective augmented reality magic lens from gradients}.
\newblock In {\em ACM Symposium on Virtual Reality Software and Technology}, pp. 87--96. ACM, 2014. \href{https://doi.org/10.1145/2671015.2671027}
{doi: {{%
10\hspace{.1pt}\discretionary{.}{%
}{.}\hspace{.4pt}1145\discretionary{/}{%
}{/}2671015\hspace{.1pt}\discretionary{.}{%
}{.}\hspace{.4pt}2671027}}}


\bibitem{Baricevic2012}
D.~Baričević, C.~Lee, M.~Turk, T.~Höllerer, and D.~A. Bowman.
\newblock A hand-held ar magic lens with user-perspective rendering.
\newblock In {\em 2012 IEEE International Symposium on Mixed and Augmented Reality (ISMAR)}, pp. 197--206, 2012. \href{https://doi.org/10.1109/ISMAR.2012.6402557}
{doi: {{%
10\hspace{.1pt}\discretionary{.}{%
}{.}\hspace{.4pt}1109\discretionary{/}{%
}{/}ISMAR\hspace{.1pt}\discretionary{.}{%
}{.}\hspace{.4pt}2012\hspace{.1pt}\discretionary{.}{%
}{.}\hspace{.4pt}6402557}}}


\bibitem{Bhattarai2020Firefighting}
M.~Bhattarai, A.~R. Jensen-Curtis, and M.~Martínez-Ramón.
\newblock An embedded deep learning system for augmented reality in firefighting applications.
\newblock In {\em 2020 19th IEEE International Conference on Machine Learning and Applications (ICMLA)}, pp. 1224--1230, 2020. \href{https://doi.org/10.1109/ICMLA51294.2020.00193}
{doi: {{%
10\hspace{.1pt}\discretionary{.}{%
}{.}\hspace{.4pt}1109\discretionary{/}{%
}{/}ICMLA51294\hspace{.1pt}\discretionary{.}{%
}{.}\hspace{.4pt}2020\hspace{.1pt}\discretionary{.}{%
}{.}\hspace{.4pt}00193}}}


\bibitem{Borsoi2018}
R.~A. Borsoi and G.~H. Costa.
\newblock {On the Performance and Implementation of Parallax Free Video See-Through Displays}.
\newblock {\em IEEE Transactions on Visualization and Computer Graphics}, 24(6):2011--2022, 2018. \href{https://doi.org/10.1109/TVCG.2017.2705184}
{doi: {{%
10\hspace{.1pt}\discretionary{.}{%
}{.}\hspace{.4pt}1109\discretionary{/}{%
}{/}TVCG\hspace{.1pt}\discretionary{.}{%
}{.}\hspace{.4pt}2017\hspace{.1pt}\discretionary{.}{%
}{.}\hspace{.4pt}2705184}}}


\bibitem{Burova2020}
A.~Burova, J.~M\"{a}kel\"{a}, J.~Hakulinen, T.~Keskinen, H.~Heinonen, S.~Siltanen, and M.~Turunen.
\newblock Utilizing vr and gaze tracking to develop ar solutions for industrial maintenance.
\newblock In {\em Proceedings of the 2020 CHI Conference on Human Factors in Computing Systems}, CHI '20,  13 pages, p. 1–13. Association for Computing Machinery, New York, NY, USA, 2020. \href{https://doi.org/10.1145/3313831.3376405}
{doi: {{%
10\hspace{.1pt}\discretionary{.}{%
}{.}\hspace{.4pt}1145\discretionary{/}{%
}{/}3313831\hspace{.1pt}\discretionary{.}{%
}{.}\hspace{.4pt}3376405}}}


\bibitem{Cao2023}
J.~Cao, K.-Y. Lam, L.-H. Lee, X.~Liu, P.~Hui, and X.~Su.
\newblock Mobile augmented reality: User interfaces, frameworks, and intelligence.
\newblock {\em ACM Comput. Surv.}, 55(9),  article no. 189,  36 pages, Jan. 2023. \href{https://doi.org/10.1145/3557999}
{doi: {{%
10\hspace{.1pt}\discretionary{.}{%
}{.}\hspace{.4pt}1145\discretionary{/}{%
}{/}3557999}}}


\bibitem{10.1145/3384540}
G.~Chaurasia, A.~Nieuwoudt, A.-E. Ichim, R.~Szeliski, and A.~Sorkine-Hornung.
\newblock Passthrough+: Real-time stereoscopic view synthesis for mobile mixed reality.
\newblock {\em Proc. ACM Comput. Graph. Interact. Tech.}, 3(1),  article no. 7,  17 pages, May 2020. \href{https://doi.org/10.1145/3384540}
{doi: {{%
10\hspace{.1pt}\discretionary{.}{%
}{.}\hspace{.4pt}1145\discretionary{/}{%
}{/}3384540}}}


\bibitem{chemaly2025mind}
T.~E. Chemaly, M.~Goyal, T.~Duan, V.~Phadnis, S.~Khattar, B.~Vlaskamp, A.~Kulshrestha, E.~L. Turner, A.~Purohit, G.~Neiswander, et~al.
\newblock Mind the gap: Geometry aware passthrough mitigates cybersickness.
\newblock {\em arXiv preprint arXiv:2502.11497}, 2025.

\bibitem{Cheng2021}
Y.~Cheng, Y.~Yan, X.~Yi, Y.~Shi, and D.~Lindlbauer.
\newblock Semanticadapt: Optimization-based adaptation of mixed reality layouts leveraging virtual-physical semantic connections.
\newblock In {\em The 34th Annual ACM Symposium on User Interface Software and Technology}, UIST '21,  16 pages, p. 282–297. Association for Computing Machinery, New York, NY, USA, 2021. \href{https://doi.org/10.1145/3472749.3474750}
{doi: {{%
10\hspace{.1pt}\discretionary{.}{%
}{.}\hspace{.4pt}1145\discretionary{/}{%
}{/}3472749\hspace{.1pt}\discretionary{.}{%
}{.}\hspace{.4pt}3474750}}}


\bibitem{10.1145/3654777.3676379}
M.~D. Dogan, E.~J. Gonzalez, K.~Ahuja, R.~Du, A.~Cola\c{c}o, J.~Lee, M.~Gonzalez-Franco, and D.~Kim.
\newblock Augmented object intelligence with xr-objects.
\newblock In {\em Proceedings of the 37th Annual ACM Symposium on User Interface Software and Technology}, UIST '24,  article no. 19,  15 pages. Association for Computing Machinery, New York, NY, USA, 2024. \href{https://doi.org/10.1145/3654777.3676379}
{doi: {{%
10\hspace{.1pt}\discretionary{.}{%
}{.}\hspace{.4pt}1145\discretionary{/}{%
}{/}3654777\hspace{.1pt}\discretionary{.}{%
}{.}\hspace{.4pt}3676379}}}


\bibitem{Elkin2021}
L.~A. Elkin, M.~Kay, J.~J. Higgins, and J.~O. Wobbrock.
\newblock An aligned rank transform procedure for multifactor contrast tests.
\newblock In {\em The 34th Annual ACM Symposium on User Interface Software and Technology}, UIST '21,  15 pages, p. 754–768, 2021. \href{https://doi.org/10.1145/3472749.3474784}
{doi: {{%
10\hspace{.1pt}\discretionary{.}{%
}{.}\hspace{.4pt}1145\discretionary{/}{%
}{/}3472749\hspace{.1pt}\discretionary{.}{%
}{.}\hspace{.4pt}3474784}}}


\bibitem{Emsenhuber2022}
G.~Emsenhuber, M.~Domhardt, T.~Langlotz, D.~Kalkofen, and M.~Tatzgern.
\newblock Towards eye-perspective rendering for optical see-through head-mounted displays.
\newblock In {\em 2022 IEEE Conference on Virtual Reality and 3D User Interfaces Abstracts and Workshops (VRW)}, pp. 640--641, 2022. \href{https://doi.org/10.1109/VRW55335.2022.00171}
{doi: {{%
10\hspace{.1pt}\discretionary{.}{%
}{.}\hspace{.4pt}1109\discretionary{/}{%
}{/}VRW55335\hspace{.1pt}\discretionary{.}{%
}{.}\hspace{.4pt}2022\hspace{.1pt}\discretionary{.}{%
}{.}\hspace{.4pt}00171}}}


\bibitem{Emsenhuber2023}
G.~Emsenhuber, T.~Langlotz, D.~Kalkofen, J.~Sutton, and M.~Tatzgern.
\newblock Eye-perspective view management for optical see-through head-mounted displays.
\newblock In {\em Proceedings of the 2023 CHI Conference on Human Factors in Computing Systems}, CHI '23,  article no. 707,  16 pages. Association for Computing Machinery, New York, NY, USA, 2023. \href{https://doi.org/10.1145/3544548.3581059}
{doi: {{%
10\hspace{.1pt}\discretionary{.}{%
}{.}\hspace{.4pt}1145\discretionary{/}{%
}{/}3544548\hspace{.1pt}\discretionary{.}{%
}{.}\hspace{.4pt}3581059}}}


\bibitem{Eswaran2023ARAssembly}
M.~Eswaran, A.~K. Gulivindala, A.~K. Inkulu, and M.~{Raju Bahubalendruni}.
\newblock Augmented reality-based guidance in product assembly and maintenance/repair perspective: A state of the art review on challenges and opportunities.
\newblock {\em Expert Systems with Applications}, 213:118983, 2023. \href{https://doi.org/10.1016/j.eswa.2022.118983}
{doi: {{%
10\hspace{.1pt}\discretionary{.}{%
}{.}\hspace{.4pt}1016\discretionary{/}{%
}{/}j\hspace{.1pt}\discretionary{.}{%
}{.}\hspace{.4pt}eswa\hspace{.1pt}\discretionary{.}{%
}{.}\hspace{.4pt}2022\hspace{.1pt}\discretionary{.}{%
}{.}\hspace{.4pt}118983}}}


\bibitem{Gabbard2007}
J.~L. Gabbard, J.~E. Swan, D.~Hix, {Si-Jung Kim}, and G.~Fitch.
\newblock {Active Text Drawing Styles for Outdoor Augmented Reality: A User-Based Study and Design Implications}.
\newblock In {\em IEEE Virtual Reality}, pp. 35--42. IEEE, mar 2007. \href{https://doi.org/10.1109/VR.2007.352461}
{doi: {{%
10\hspace{.1pt}\discretionary{.}{%
}{.}\hspace{.4pt}1109\discretionary{/}{%
}{/}VR\hspace{.1pt}\discretionary{.}{%
}{.}\hspace{.4pt}2007\hspace{.1pt}\discretionary{.}{%
}{.}\hspace{.4pt}352461}}}


\bibitem{Ghasemi2022DeepLearning}
Y.~Ghasemi, H.~Jeong, S.~H. Choi, K.-B. Park, and J.~Y. Lee.
\newblock Deep learning-based object detection in augmented reality: A systematic review.
\newblock {\em Computers in Industry}, 139:103661, 2022. \href{https://doi.org/10.1016/j.compind.2022.103661}
{doi: {{%
10\hspace{.1pt}\discretionary{.}{%
}{.}\hspace{.4pt}1016\discretionary{/}{%
}{/}j\hspace{.1pt}\discretionary{.}{%
}{.}\hspace{.4pt}compind\hspace{.1pt}\discretionary{.}{%
}{.}\hspace{.4pt}2022\hspace{.1pt}\discretionary{.}{%
}{.}\hspace{.4pt}103661}}}


\bibitem{Gsaxner2024}
C.~Gsaxner, S.~Mori, D.~Schmalstieg, J.~Egger, G.~Paar, W.~Bailer, and D.~Kalkofen.
\newblock Deepdr: Deep structure-aware rgb-d inpainting for diminished reality.
\newblock In {\em 2024 International Conference on 3D Vision (3DV)}, pp. 750--760, 2024. \href{https://doi.org/10.1109/3DV62453.2024.00037}
{doi: {{%
10\hspace{.1pt}\discretionary{.}{%
}{.}\hspace{.4pt}1109\discretionary{/}{%
}{/}3DV62453\hspace{.1pt}\discretionary{.}{%
}{.}\hspace{.4pt}2024\hspace{.1pt}\discretionary{.}{%
}{.}\hspace{.4pt}00037}}}


\bibitem{Gu2024}
Q.~Gu, A.~Kuwajerwala, S.~Morin, K.~M. Jatavallabhula, B.~Sen, A.~Agarwal, C.~Rivera, W.~Paul, K.~Ellis, R.~Chellappa, C.~Gan, C.~M. de~Melo, J.~B. Tenenbaum, A.~Torralba, F.~Shkurti, and L.~Paull.
\newblock Conceptgraphs: Open-vocabulary 3d scene graphs for perception and planning.
\newblock In {\em 2024 IEEE International Conference on Robotics and Automation (ICRA)}, pp. 5021--5028, 2024. \href{https://doi.org/10.1109/ICRA57147.2024.10610243}
{doi: {{%
10\hspace{.1pt}\discretionary{.}{%
}{.}\hspace{.4pt}1109\discretionary{/}{%
}{/}ICRA57147\hspace{.1pt}\discretionary{.}{%
}{.}\hspace{.4pt}2024\hspace{.1pt}\discretionary{.}{%
}{.}\hspace{.4pt}10610243}}}


\bibitem{10.1145/3610548.3618134}
P.~Guan, E.~Penner, J.~Hegland, B.~Letham, and D.~Lanman.
\newblock Perceptual requirements for world-locked rendering in ar and vr.
\newblock In {\em SIGGRAPH Asia 2023 Conference Papers}, SA '23,  article no. 35,  10 pages. Association for Computing Machinery, New York, NY, USA, 2023. \href{https://doi.org/10.1145/3610548.3618134}
{doi: {{%
10\hspace{.1pt}\discretionary{.}{%
}{.}\hspace{.4pt}1145\discretionary{/}{%
}{/}3610548\hspace{.1pt}\discretionary{.}{%
}{.}\hspace{.4pt}3618134}}}


\bibitem{HART1988}
S.~G. Hart and L.~E. Staveland.
\newblock {Development of NASA-TLX (Task Load Index): Results of Empirical and Theoretical Research}.
\newblock In P.~A. Hancock and N.~Meshkati, eds., {\em Human Mental Workload}, vol.~52 of {\em Advances in Psychology}, pp. 139--183. North-Holland, 1988.

\bibitem{Hwang2014GoogleGlass}
A.~D. Hwang and E.~Peli.
\newblock An augmented-reality edge enhancement application for google glass.
\newblock {\em Optometry and Vision Science}, 91(8), 2014.

\bibitem{Jia2021}
J.~Jia, S.~Elezovikj, H.~Fan, S.~Yang, J.~Liu, W.~Guo, C.~C. Tan, and H.~Ling.
\newblock {Semantic-aware label placement for augmented reality in street view}.
\newblock {\em Visual Computer}, 37(7):1805--1819, 2021. \href{https://doi.org/10.1007/s00371-020-01939-w}
{doi: {{%
10\hspace{.1pt}\discretionary{.}{%
}{.}\hspace{.4pt}1007\discretionary{/}{%
}{/}s00371\discretionary{%
}{-}{-}020\discretionary{%
}{-}{-}01939\discretionary{%
}{-}{-}w}}}


\bibitem{Kim2024}
S.~K. Kim and M.~Y. Kim.
\newblock Deep learning-based calibration method for an augmented reality surgical navigation system without head-mounted optical markers.
\newblock In {\em 2024 24th International Conference on Control, Automation and Systems (ICCAS)}, pp. 453--458, 2024. \href{https://doi.org/10.23919/ICCAS63016.2024.10773365}
{doi: {{%
10\hspace{.1pt}\discretionary{.}{%
}{.}\hspace{.4pt}23919\discretionary{/}{%
}{/}ICCAS63016\hspace{.1pt}\discretionary{.}{%
}{.}\hspace{.4pt}2024\hspace{.1pt}\discretionary{.}{%
}{.}\hspace{.4pt}10773365}}}


\bibitem{10.1145/3588432.3591534}
G.~Kuo, E.~Penner, S.~Moczydlowski, A.~Ching, D.~Lanman, and N.~Matsuda.
\newblock Perspective-correct vr passthrough without reprojection.
\newblock In {\em ACM SIGGRAPH 2023 Conference Proceedings}, SIGGRAPH '23,  article no. 15,  9 pages. Association for Computing Machinery, New York, NY, USA, 2023. \href{https://doi.org/10.1145/3588432.3591534}
{doi: {{%
10\hspace{.1pt}\discretionary{.}{%
}{.}\hspace{.4pt}1145\discretionary{/}{%
}{/}3588432\hspace{.1pt}\discretionary{.}{%
}{.}\hspace{.4pt}3591534}}}


\bibitem{Lai2020}
Z.-H. Lai, W.~Tao, M.~C. Leu, and Z.~Yin.
\newblock Smart augmented reality instructional system for mechanical assembly towards worker-centered intelligent manufacturing.
\newblock {\em Journal of Manufacturing Systems}, 55:69--81, 2020. \href{https://doi.org/10.1016/j.jmsy.2020.02.010}
{doi: {{%
10\hspace{.1pt}\discretionary{.}{%
}{.}\hspace{.4pt}1016\discretionary{/}{%
}{/}j\hspace{.1pt}\discretionary{.}{%
}{.}\hspace{.4pt}jmsy\hspace{.1pt}\discretionary{.}{%
}{.}\hspace{.4pt}2020\hspace{.1pt}\discretionary{.}{%
}{.}\hspace{.4pt}02\hspace{.1pt}\discretionary{.}{%
}{.}\hspace{.4pt}010}}}


\bibitem{Langlotz2016}
T.~Langlotz, M.~Cook, and H.~Regenbrecht.
\newblock {Real-time radiometric compensation for optical see-through head-mounted displays}.
\newblock {\em IEEE Transactions on Visualization and Computer Graphics}, 22(11):2385--2394, 2016. \href{https://doi.org/10.1109/TVCG.2016.2593781}
{doi: {{%
10\hspace{.1pt}\discretionary{.}{%
}{.}\hspace{.4pt}1109\discretionary{/}{%
}{/}TVCG\hspace{.1pt}\discretionary{.}{%
}{.}\hspace{.4pt}2016\hspace{.1pt}\discretionary{.}{%
}{.}\hspace{.4pt}2593781}}}


\bibitem{Langlotz2024}
T.~Langlotz, J.~Sutton, and H.~Regenbrecht.
\newblock A design space for vision augmentations and augmented human perception using digital eyewear.
\newblock In {\em Proceedings of the 2024 CHI Conference on Human Factors in Computing Systems}, CHI '24,  article no. 966,  16 pages. Association for Computing Machinery, New York, NY, USA, 2024. \href{https://doi.org/10.1145/3613904.3642380}
{doi: {{%
10\hspace{.1pt}\discretionary{.}{%
}{.}\hspace{.4pt}1145\discretionary{/}{%
}{/}3613904\hspace{.1pt}\discretionary{.}{%
}{.}\hspace{.4pt}3642380}}}


\bibitem{Langlotz2018}
T.~Langlotz, J.~Sutton, S.~Zollmann, Y.~Itoh, and H.~Regenbrecht.
\newblock Chromaglasses: Computational glasses for compensating colour blindness.
\newblock In {\em Proceedings of the 2018 CHI Conference on Human Factors in Computing Systems}, CHI '18,  12 pages, p. 1–12. Association for Computing Machinery, New York, NY, USA, 2018. \href{https://doi.org/10.1145/3173574.3173964}
{doi: {{%
10\hspace{.1pt}\discretionary{.}{%
}{.}\hspace{.4pt}1145\discretionary{/}{%
}{/}3173574\hspace{.1pt}\discretionary{.}{%
}{.}\hspace{.4pt}3173964}}}


\bibitem{Liu2024HazardDetection}
J.~Liu, A.~S. Rao, F.~Ke, T.~Dwyer, B.~Tag, and P.~D. Haghighi.
\newblock Ar-facilitated safety inspection and fall hazard detection on construction sites.
\newblock In {\em 2024 IEEE International Symposium on Mixed and Augmented Reality Adjunct (ISMAR-Adjunct)}, pp. 12--14, 2024. \href{https://doi.org/10.1109/ISMAR-Adjunct64951.2024.00010}
{doi: {{%
10\hspace{.1pt}\discretionary{.}{%
}{.}\hspace{.4pt}1109\discretionary{/}{%
}{/}ISMAR\discretionary{%
}{-}{-}Adjunct64951\hspace{.1pt}\discretionary{.}{%
}{.}\hspace{.4pt}2024\hspace{.1pt}\discretionary{.}{%
}{.}\hspace{.4pt}00010}}}


\bibitem{Liu2019}
L.~Liu, H.~Li, and M.~Gruteser.
\newblock Edge assisted real-time object detection for mobile augmented reality.
\newblock In {\em The 25th Annual International Conference on Mobile Computing and Networking}, MobiCom '19,  article no. 25,  16 pages. Association for Computing Machinery, New York, NY, USA, 2019. \href{https://doi.org/10.1145/3300061.3300116}
{doi: {{%
10\hspace{.1pt}\discretionary{.}{%
}{.}\hspace{.4pt}1145\discretionary{/}{%
}{/}3300061\hspace{.1pt}\discretionary{.}{%
}{.}\hspace{.4pt}3300116}}}


\bibitem{Lutnyk2024}
L.~Lutnyk, K.~G. Kim, A.~Sarbach, P.~Kiefer, R.~Häusler, and M.~R. and.
\newblock Context-sensitive augmented reality assistance in the cockpit.
\newblock {\em International Journal of Human–Computer Interaction}, 0(0):1--20, 2024. \href{https://doi.org/10.1080/10447318.2024.2440976}
{doi: {{%
10\hspace{.1pt}\discretionary{.}{%
}{.}\hspace{.4pt}1080\discretionary{/}{%
}{/}10447318\hspace{.1pt}\discretionary{.}{%
}{.}\hspace{.4pt}2024\hspace{.1pt}\discretionary{.}{%
}{.}\hspace{.4pt}2440976}}}


\bibitem{Mardanbegi2019}
D.~Mardanbegi, T.~Langlotz, and H.~Gellersen.
\newblock Resolving target ambiguity in 3d gaze interaction through vor depth estimation.
\newblock In {\em Proceedings of the 2019 CHI Conference on Human Factors in Computing Systems}, CHI '19,  12 pages, p. 1–12. Association for Computing Machinery, New York, NY, USA, 2019. \href{https://doi.org/10.1145/3290605.3300842}
{doi: {{%
10\hspace{.1pt}\discretionary{.}{%
}{.}\hspace{.4pt}1145\discretionary{/}{%
}{/}3290605\hspace{.1pt}\discretionary{.}{%
}{.}\hspace{.4pt}3300842}}}


\bibitem{Matsuda2013}
Y.~Matsuda, F.~Shibata, A.~Kimura, and H.~Tamura.
\newblock {Poster: Creating a user-specific perspective view for mobile mixed reality systems on smartphones}.
\newblock In {\em IEEE Symposium on 3D User Interface}, pp. 157--158. IEEE, 2013. \href{https://doi.org/10.1109/3DUI.2013.6550226}
{doi: {{%
10\hspace{.1pt}\discretionary{.}{%
}{.}\hspace{.4pt}1109\discretionary{/}{%
}{/}3DUI\hspace{.1pt}\discretionary{.}{%
}{.}\hspace{.4pt}2013\hspace{.1pt}\discretionary{.}{%
}{.}\hspace{.4pt}6550226}}}


\bibitem{Miles1930}
W.~R. Miles.
\newblock Ocular dominance in human adults.
\newblock {\em The Journal of General Psychology}, 3(3):412--430, 1930. \href{https://doi.org/10.1080/00221309.1930.9918218}
{doi: {{%
10\hspace{.1pt}\discretionary{.}{%
}{.}\hspace{.4pt}1080\discretionary{/}{%
}{/}00221309\hspace{.1pt}\discretionary{.}{%
}{.}\hspace{.4pt}1930\hspace{.1pt}\discretionary{.}{%
}{.}\hspace{.4pt}9918218}}}


\bibitem{Htike2021}
H.~Min~Htike, T.~H.~Margrain, Y.-K. Lai, and P.~Eslambolchilar.
\newblock Augmented reality glasses as an orientation and mobility aid for people with low vision: a feasibility study of experiences and requirements.
\newblock In {\em Proceedings of the 2021 CHI Conference on Human Factors in Computing Systems}, CHI '21,  article no. 729,  15 pages. Association for Computing Machinery, New York, NY, USA, 2021. \href{https://doi.org/10.1145/3411764.3445327}
{doi: {{%
10\hspace{.1pt}\discretionary{.}{%
}{.}\hspace{.4pt}1145\discretionary{/}{%
}{/}3411764\hspace{.1pt}\discretionary{.}{%
}{.}\hspace{.4pt}3445327}}}


\bibitem{7893336}
P.~Mohr, M.~Tatzgern, J.~Grubert, D.~Schmalstieg, and D.~Kalkofen.
\newblock Adaptive user perspective rendering for handheld augmented reality.
\newblock In {\em 2017 IEEE Symposium on 3D User Interfaces (3DUI)}, pp. 176--181, 2017. \href{https://doi.org/10.1109/3DUI.2017.7893336}
{doi: {{%
10\hspace{.1pt}\discretionary{.}{%
}{.}\hspace{.4pt}1109\discretionary{/}{%
}{/}3DUI\hspace{.1pt}\discretionary{.}{%
}{.}\hspace{.4pt}2017\hspace{.1pt}\discretionary{.}{%
}{.}\hspace{.4pt}7893336}}}


\bibitem{Mooser2007}
J.~Mooser, S.~You, and U.~Neumann.
\newblock Real-time object tracking for augmented reality combining graph cuts and optical flow.
\newblock In {\em 2007 6th IEEE and ACM International Symposium on Mixed and Augmented Reality}, pp. 145--152, 2007. \href{https://doi.org/10.1109/ISMAR.2007.4538839}
{doi: {{%
10\hspace{.1pt}\discretionary{.}{%
}{.}\hspace{.4pt}1109\discretionary{/}{%
}{/}ISMAR\hspace{.1pt}\discretionary{.}{%
}{.}\hspace{.4pt}2007\hspace{.1pt}\discretionary{.}{%
}{.}\hspace{.4pt}4538839}}}


\bibitem{WHO_Blindness_2023}
W.~H. Organization.
\newblock Blindness and visual impairment, 2023.
\newblock Accessed: 2025-03-24.

\bibitem{Rajapaksha2024}
U.~Rajapaksha, F.~Sohel, H.~Laga, D.~Diepeveen, and M.~Bennamoun.
\newblock Deep learning-based depth estimation methods from monocular image and videos: A comprehensive survey.
\newblock {\em ACM Comput. Surv.}, 56(12),  article no. 315,  51 pages, Oct. 2024. \href{https://doi.org/10.1145/3677327}
{doi: {{%
10\hspace{.1pt}\discretionary{.}{%
}{.}\hspace{.4pt}1145\discretionary{/}{%
}{/}3677327}}}


\bibitem{Sutton2022}
J.~Sutton, T.~Langlotz, and A.~Plopski.
\newblock Seeing colours: Addressing colour vision deficiency with vision augmentations using computational glasses.
\newblock {\em ACM Trans. Comput.-Hum. Interact.}, 29(3),  article no. 26,  53 pages, Jan. 2022. \href{https://doi.org/10.1145/3486899}
{doi: {{%
10\hspace{.1pt}\discretionary{.}{%
}{.}\hspace{.4pt}1145\discretionary{/}{%
}{/}3486899}}}


\bibitem{sutton2022look}
J.~Sutton, T.~Langlotz, A.~Plopski, S.~Zollmann, Y.~Itoh, and H.~Regenbrecht.
\newblock Look over there! investigating saliency modulation for visual guidance with augmented reality glasses.
\newblock In {\em Proceedings of the 35th Annual ACM Symposium on User Interface Software and Technology}, pp. 1--15, 2022.

\bibitem{Teruggi22HoloLens}
S.~Teruggi and F.~Fassi.
\newblock Hololens 2 spatial mapping capabilities in vast monumental heritage environments.
\newblock {\em The International Archives of the Photogrammetry, Remote Sensing and Spatial Information Sciences}, XLVI-2/W1-2022:489--496, 2022. \href{https://doi.org/10.5194/isprs-archives-XLVI-2-W1-2022-489-2022}
{doi: {{%
10\hspace{.1pt}\discretionary{.}{%
}{.}\hspace{.4pt}5194\discretionary{/}{%
}{/}isprs\discretionary{%
}{-}{-}archives\discretionary{%
}{-}{-}XLVI\discretionary{%
}{-}{-}2\discretionary{%
}{-}{-}W1\discretionary{%
}{-}{-}2022\discretionary{%
}{-}{-}489\discretionary{%
}{-}{-}2022}}}


\bibitem{Toyama2013}
T.~Toyama, W.~Suzuki, A.~Dengel, and K.~Kise.
\newblock User attention oriented augmented reality on documents with document dependent dynamic overlay.
\newblock In {\em 2013 IEEE International Symposium on Mixed and Augmented Reality (ISMAR)}, pp. 299--300, 2013. \href{https://doi.org/10.1109/ISMAR.2013.6671814}
{doi: {{%
10\hspace{.1pt}\discretionary{.}{%
}{.}\hspace{.4pt}1109\discretionary{/}{%
}{/}ISMAR\hspace{.1pt}\discretionary{.}{%
}{.}\hspace{.4pt}2013\hspace{.1pt}\discretionary{.}{%
}{.}\hspace{.4pt}6671814}}}


\bibitem{Wang22}
Z.~Wang, Y.~Zhao, and F.~Lu.
\newblock Gaze-vergence-controlled see-through vision in augmented reality.
\newblock {\em IEEE Transactions on Visualization and Computer Graphics}, 28(11):3843--3853, 2022. \href{https://doi.org/10.1109/TVCG.2022.3203110}
{doi: {{%
10\hspace{.1pt}\discretionary{.}{%
}{.}\hspace{.4pt}1109\discretionary{/}{%
}{/}TVCG\hspace{.1pt}\discretionary{.}{%
}{.}\hspace{.4pt}2022\hspace{.1pt}\discretionary{.}{%
}{.}\hspace{.4pt}3203110}}}


\bibitem{Weinmann2021}
M.~Weinmann, S.~Wursthorn, M.~Weinmann, and P.~Hübner.
\newblock Efficient {3D} {Mapping} and {Modelling} of {Indoor} {Scenes} with the {Microsoft} {HoloLens}: {A} {Survey}.
\newblock {\em PFG – Journal of Photogrammetry, Remote Sensing and Geoinformation Science}, 89(4):319--333, Aug. 2021. \href{https://doi.org/10.1007/s41064-021-00163-y}
{doi: {{%
10\hspace{.1pt}\discretionary{.}{%
}{.}\hspace{.4pt}1007\discretionary{/}{%
}{/}s41064\discretionary{%
}{-}{-}021\discretionary{%
}{-}{-}00163\discretionary{%
}{-}{-}y}}}


\bibitem{Wobbrock2011}
J.~O. Wobbrock, L.~Findlater, D.~Gergle, and J.~J. Higgins.
\newblock The aligned rank transform for nonparametric factorial analyses using only anova procedures.
\newblock In {\em Proceedings of the SIGCHI Conference on Human Factors in Computing Systems}, CHI '11,  4 pages, p. 143–146, 2011. \href{https://doi.org/10.1145/1978942.1978963}
{doi: {{%
10\hspace{.1pt}\discretionary{.}{%
}{.}\hspace{.4pt}1145\discretionary{/}{%
}{/}1978942\hspace{.1pt}\discretionary{.}{%
}{.}\hspace{.4pt}1978963}}}


\bibitem{10.1145/3528233.3530701}
L.~Xiao, S.~Nouri, J.~Hegland, A.~G. Garcia, and D.~Lanman.
\newblock Neuralpassthrough: Learned real-time view synthesis for vr.
\newblock In {\em ACM SIGGRAPH 2022 Conference Proceedings}, SIGGRAPH '22,  article no. 40,  9 pages. Association for Computing Machinery, New York, NY, USA, 2022. \href{https://doi.org/10.1145/3528233.3530701}
{doi: {{%
10\hspace{.1pt}\discretionary{.}{%
}{.}\hspace{.4pt}1145\discretionary{/}{%
}{/}3528233\hspace{.1pt}\discretionary{.}{%
}{.}\hspace{.4pt}3530701}}}


\bibitem{Xu2024}
C.~Xu, R.~Kumaran, N.~Stier, K.~Yu, and T.~Höllerer.
\newblock Multimodal 3d fusion and in-situ learning for spatially aware ai.
\newblock In {\em 2024 IEEE International Symposium on Mixed and Augmented Reality (ISMAR)}, pp. 485--494, 2024. \href{https://doi.org/10.1109/ISMAR62088.2024.00063}
{doi: {{%
10\hspace{.1pt}\discretionary{.}{%
}{.}\hspace{.4pt}1109\discretionary{/}{%
}{/}ISMAR62088\hspace{.1pt}\discretionary{.}{%
}{.}\hspace{.4pt}2024\hspace{.1pt}\discretionary{.}{%
}{.}\hspace{.4pt}00063}}}


\bibitem{Younis2019}
O.~Younis, W.~Al-Nuaimy, F.~Rowe, and M.~H. Alomari.
\newblock A smart context-aware hazard attention system to help people with peripheral vision loss.
\newblock {\em Sensors}, 19(7), 2019. \href{https://doi.org/10.3390/s19071630}
{doi: {{%
10\hspace{.1pt}\discretionary{.}{%
}{.}\hspace{.4pt}3390\discretionary{/}{%
}{/}s19071630}}}


\bibitem{Yu2018}
X.~Yu, G.~Yang, S.~Jones, and J.~Saniie.
\newblock Ar marker aided obstacle localization system for assisting visually impaired.
\newblock In {\em 2018 IEEE International Conference on Electro/Information Technology (EIT)}, pp. 0271--0276, 2018. \href{https://doi.org/10.1109/EIT.2018.8500166}
{doi: {{%
10\hspace{.1pt}\discretionary{.}{%
}{.}\hspace{.4pt}1109\discretionary{/}{%
}{/}EIT\hspace{.1pt}\discretionary{.}{%
}{.}\hspace{.4pt}2018\hspace{.1pt}\discretionary{.}{%
}{.}\hspace{.4pt}8500166}}}


\bibitem{Zhu2023}
Z.~Zhu, J.~Li, Y.~Tang, K.~Go, M.~Toyoura, K.~Kashiwagi, I.~Fujishiro, and X.~Mao.
\newblock Cc-glasses: Color communication support for people with color vision deficiency using augmented reality and deep learning.
\newblock In {\em Proceedings of the Augmented Humans International Conference 2023}, AHs '23,  10 pages, p. 190–199. Association for Computing Machinery, New York, NY, USA, 2023. \href{https://doi.org/10.1145/3582700.3582707}
{doi: {{%
10\hspace{.1pt}\discretionary{.}{%
}{.}\hspace{.4pt}1145\discretionary{/}{%
}{/}3582700\hspace{.1pt}\discretionary{.}{%
}{.}\hspace{.4pt}3582707}}}


\bibitem{Lysakowski2023Yolo}
M.~Łysakowski, K.~Żywanowski, A.~Banaszczyk, M.~R. Nowicki, P.~Skrzypczyński, and S.~K. Tadeja.
\newblock Real-time onboard object detection for augmented reality: Enhancing head-mounted display with yolov8.
\newblock In {\em 2023 IEEE International Conference on Edge Computing and Communications (EDGE)}, pp. 364--371, 2023. \href{https://doi.org/10.1109/EDGE60047.2023.00059}
{doi: {{%
10\hspace{.1pt}\discretionary{.}{%
}{.}\hspace{.4pt}1109\discretionary{/}{%
}{/}EDGE60047\hspace{.1pt}\discretionary{.}{%
}{.}\hspace{.4pt}2023\hspace{.1pt}\discretionary{.}{%
}{.}\hspace{.4pt}00059}}}


\end{thebibliography}
\balance
\end{document}